\documentclass{jltp}
  
\usepackage{graphicx}
\usepackage{amssymb}
\usepackage{amsfonts}
\usepackage{bm}

\newcommand{\text}[1]{\mbox{#1}}

\newcommand{\rem}[1]{}

\begin{document}

\title{Low-Energy Scales and Temperature-Dependent Photoemission
of Heavy Fermions\thanks{Dedicated to Peter 
W\"olfle on the occasion of his 60th birthday.}}

\author{T. A. Costi and N. Manini$^\dagger$}

\address
{European Synchrotron Radiation Facility, BP 220,\\
F-38043 Grenoble Cedex, France\\
$^\dagger$Dipartimento di Fisica, Universit\`a di Milano,
Via Celoria 16 - 20133 Milano - Italy\\ and INFM, 
Unit\`a di Milano, Milano, Italy
}

\runninghead{T. A. Costi and N. Manini}{Temperature-Dependent 
Photoemission of Heavy Fermions}

\maketitle
\begin{abstract}
We solve the $S=1/2$ Kondo lattice model within the dynamical mean field
theory.
Detailed predictions are made for the dependence of the lattice
Kondo resonance and the conduction electron spectral density
on temperature and band filling, $n_{c}$.
Two low-energy scales are identified in the spectra: a renormalized 
hybridization pseudogap scale $T^{*}$, which correlates with 
the single-ion Kondo scale, and a lattice Kondo scale $T_{0}\ll T^{*}$,
which acts as the Fermi-liquid coherence scale. The lattice Kondo
resonance is split into a main branch, which is pinned at the Fermi level, 
and whose width is set by $T_{0}$, and an upper branch at 
$\omega\approx T^{*}$. The weight of the upper branch decreases 
rapidly away from $n_{c}=1$ and vanishes for $n_{c}\lesssim 0.7$ 
(however, the pseudogap in the conduction electron spectral density persists
for all $n_{c}$). 
On increasing temperature, we find that the lattice Kondo resonance 
vanishes on a temperature scale of order 
$10\,T_{0}$, the same 
scale over which the single-ion Kondo resonance vanishes in impurity 
model calculations.
In contrast to impurity model calculations, however, we find that 
the position of the lattice Kondo resonance depends strongly on 
temperature.
The results are used to make predictions on the temperature
dependence of the low-energy photoemission
spectrum of metallic heavy fermions and doped Kondo insulators. 
We compare our results
for the photoemission spectra with available high-resolution measurements 
on YbInCu$_4$ and YbAgCu$_4$. The loss in intensity with increasing
temperature, and the asymmetric lineshape of the low-energy
spectra are well accounted for by our simplified $S=1/2$ Kondo lattice
model. 

\end{abstract}

\section{INTRODUCTION}

Heavy fermions exhibit a number of interesting 
ground states with anomalous low-temperature properties 
\cite{kuramoto.00,hewson.93,grewe.91,wachter.94,lee.86,stewart.84}. 
These include 
paramagnetic heavy-fermion states down to the lowest temperatures 
(e.g.\ CeCu$_{6}$, CeAl$_{3}$, CeRu$_{2}$Si$_{2}$, YbAgCu$_{4}$, and 
YbInCu$_{4}$), magnetically ordered states with small ordering 
temperatures and reduced moments (e.g.\ URu$_{2}$Si$_{2}$, CeAl$_{2}$), 
and unconventional superconducting states (e.g.\ CeCu$_{2}$Si$_{2}$, 
UBe$_{13}$, UPt$_{3}$).
This anomalous behavior
is mostly attributed to the strong correlations between the electrons 
in the localized or weakly-itinerant $f$ states of the rare-earth (or 
actinide) ions.
The natural starting point for understanding this
peculiar behavior is the periodic Anderson model or, 
when charge fluctuations are negligible, its low-energy counterpart, 
the Kondo lattice model. 
These models are fairly well understood
in one dimension where extensive
calculations, based on exact diagonalization\cite{tsunetsugo.97}
and density matrix 
renormalization group\cite{shibata.99} techniques 
are available.
However, it is not clear how much of the insight gained
is relevant to real-life three-dimensional heavy fermions.
The Gutzwiller approximation, that can be applied in any 
dimension, has yielded some information on various correlated 
lattice models\cite{fazekas.99}, particularly for ground-state
properties.
The development of a non-perturbative
approximation scheme, the dynamical mean field theory (DMFT) 
\cite{metzner.89,mh.89,ohkawa.91,kotliar.92,jarrell.92}, has
recently enabled a more comprehensive investigation of correlated 
lattice models, and particularly of their dynamic properties 
\cite{georges.96}. 
In this paper, we apply this technique to the Kondo lattice model in order
to investigate (i) the relevant low-energy scales of the model and how
these appear in the single-particle and magnetic excitation spectra, and
(ii) the temperature dependence of the $f$-electron spectra for the
interpretation of the low-energy photoemission spectra (PES) of
paramagnetic heavy-fermion compounds, of
current experimental interest\cite{allen.00,arko.99,malterre.96}.

The $f$-level photoemission and inverse photoemission spectra (IPES) of heavy
fermions measure, respectively, the single-particle removal and addition
spectrum for $f$ electrons.
They provide information on the electronic structure 
of heavy fermions, including the energies of the relevant $f$-level 
configurations, their spin-orbit splittings, and the size of the 
screened Coulomb interaction between the $f$ electrons.
In addition to this information about the ``high-energy'' scales (the
smallest of these being the spin-orbit splitting $\sim
1$~eV), the $f$-level spectrum contains also information about low-energy
scales associated with correlations and lattice coherence.
Identifying these is one of the aims of the present study of 
the Kondo lattice model.

We shall not deal, here, with the ``high-energy'' part of the spectra of heavy 
fermions, which is  well accounted for
by an Anderson impurity model (AIM) approach 
\cite{allen.86,gs.83,bickers.87}. 
Instead, we shall focus our attention on the low-energy spectrum,
which is the part most strongly influenced by lattice coherence 
effects, especially at low temperatures.
These effects should emerge most clearly in the Kondo lattice model, which
addresses specifically the low-energy scale, neglecting the complications
associated with the high-energy atomic-like features.
Although single-impurity model approaches also make predictions for 
the low-energy spectrum, there is no a priori reason to expect these 
to be quantitatively or even qualitatively correct. A lattice model
calculation is therefore required to quantify the extent to which 
AIM calculations apply. 

%
%
In the absence of detailed lattice-model calculations, the experimental 
results for the low-energy spectra of heavy fermions have, nevertheless,
been discussed within the AIM approach. Two specific predictions of 
this approach concern the magnitude of the temperature 
dependence of the near-$E_{\rm F}$ feature (Kondo resonance) in 
the spectra, and the scaling of the intensity of this feature 
with the Kondo scale. Broad agreement with AIM predictions is claimed 
by some groups\cite{malterre.96,patthey.90,weibel.93,tjeng.93,garnier.97} 
whereas others\cite{arko.99,joyce.92,blyth.93} dispute 
the interpretation in terms of the AIM. The reader is referred 
to the reviews Ref.~\onlinecite{arko.99,malterre.96}, and to 
Ref.~\onlinecite{reinert.98,joyce.01,reinert.01a,moore.00}, for some
recent aspects of this controversy. In the context of this discussion,
these experiments address a very important issue, namely the temperature
dependence of the low-energy spectra and the appropriate theoretical
framework for interpreting them. We shall attempt to clarify this
issue by calculating the temperature dependence of the lattice
Kondo resonance and comparing this with the predictions of the 
impurity models
and with experimental spectra.
Experimental studies of this temperature dependence at 
temperatures both above and below the
relevant Kondo scale are technically difficult, and so far have only been
carried out on a few systems, for example on YbAgCu$_{4}$\cite{weibel.93}, 
YbAl$_{3}$\cite{tjeng.93}, and
more recently on YbInCu$_{4}$\cite{moore.00} and URu$_{2}$Si$_{2}$ 
\cite{denlinger.01}.
Most other studies have only compared spectra at two
temperatures, typically room temperature and at one low temperature
comparable to the Kondo scale: this is not sufficient for an unambiguous
testing of model predictions with experiment, but can at most indicate
agreement with predicted trends.
Hence, it is important for a detailed comparison with theory for 
more experiments of the former type to be carried out.

This paper is organized as follows: in Sect.~\ref{model+method:sec} 
we introduce the quantities of interest for the Kondo lattice model, 
and sketch the dynamical mean field method used; in 
Sect.~\ref{results:sec} we discuss the main theoretical findings; 
in Sect.~\ref{int:sec} we describe the results for the PES
and IPES intensities; Sect.~\ref{exp:sec} contains the
comparison of the theoretical spectra with actual PES data for
YbInCu$_4$ and YbAgCu$_4$; Sect.~\ref{conclusions:sec} summarizes the
main findings and problems left open by this paper; \ref{app:details}
provides details on the implementation of the DMFT to the Kondo lattice
model and an expression for the self energy; \ref{app:spectra}
defines the relevant spectral functions for this study.

\section{THE KONDO LATTICE MODEL AND THE DMFT METHOD}
\label{model+method:sec}

\subsection{The Kondo Lattice Model}
\label{model}
Consider the Kondo lattice model, given by the Hamiltonian
\begin{eqnarray}
{\cal H} &=& \sum_{\mathbf k,\sigma}\epsilon_{\mathbf k}~
c_{\mathbf k,\sigma}^{\dagger}
c_{\mathbf k,\sigma} 
+ \sum_{j}J\;{\mathbf S}_{j}\cdot {\mathbf s}_{j}
\ .
\label{eq:KLM}
\end{eqnarray}
The first term describes non-interacting itinerant
electrons with dispersion relation $\epsilon_{\mathbf k}$. 
The second term is the antiferromagnetic
exchange interaction, of strength $J$, between the localized spins
${\mathbf S}_{j}$ at lattice sites $j$ and the local conduction
electron spin density
$
{\mathbf s}_{j}=\sum_{\mu,\nu} c_{j,\mu}^{\dagger}
{\bm \sigma}_{\mu\nu}
c_{j,\nu}
$, where
$c_{j,\sigma}=N_{\rm sites}^{-1/2}\sum_{\mathbf k} c_{{\mathbf k},\sigma}
\exp(i{\mathbf k}\cdot {\mathbf R}_{j})$
destroys a local conduction electron orbital at site $j$.
This model is relevant for heavy fermions with negligible charge
fluctuations on the $f$ orbitals: it is appropriate for describing ${\rm
Ce}$ and ${\rm Yb}$ compounds having a doublet crystal-field ground state
for the $4f^{1}$ (Ce) or $4f^{13}$ (Yb) atomic configuration, acting as a
$S=1/2$ pseudospin represented in the model by the ${\mathbf S}_{j}$
operator.

For the unperturbed density of states of the lattice, $\rho_{0}(\omega)$,
we take a semi-elliptic form
\begin{equation}
\rho_{0}(\omega)=\sum_{\mathbf k}
\delta(\omega-\epsilon_{\mathbf k})={2\over \pi D^2} 
\sqrt{D^2 -\omega^2} \;,
\label{eq:semi-elliptic}
\end{equation} 
corresponding to a Bethe lattice.
The main features of the local dynamical quantities discussed below are
primarily a result of the correlations and not of this particular choice of
lattice, which is especially convenient in the DMFT.

\subsection{DMFT and Numerical Renormalization Group}
\label{method}

To treat the complicated correlation problem expressed by 
Eq.~(\ref{eq:KLM}) we resort to a standard technique developed
in the context of the infinite dimensional limit of correlated
lattice models\cite{metzner.89}, the DMFT\cite{georges.96}.
In the DMFT approximation, the solution of (\ref{eq:KLM}) reduces to
solving a quantum impurity model embedded in an effective conduction
electron medium, whose spectral density
$\tilde{\rho}(\omega)=\sum_{\mathbf k}\delta(\omega-\tilde{\epsilon}_{\mathbf
k})$ is determined self-consistently 
\cite{ohkawa.91,kotliar.92,jarrell.92}. In the present
case of the Kondo lattice model, the effective quantum impurity model is
just the $S=1/2$ Kondo model\cite{georges.92}
\begin{equation}
{\cal H}_{imp} =
\sum_{{\mathbf k},\sigma}\tilde{\epsilon}_{\mathbf k}\;
c_{{\mathbf k},\sigma}^{\dagger} c_{{\mathbf k},\sigma} 
+ J\;{\mathbf S}\cdot {\mathbf s}
\label{eq:KM} \ ,
\end{equation}
where ${\mathbf S}$ is the impurity spin and  ${\mathbf s}$ is
the spin density of the conduction electrons at the impurity site.
The effective medium is determined from the self-consistency
requirement that the local conduction-electron Green's function
of the effective impurity model $G_{c}(\omega,T)$ (which
depends implicitly on $\tilde{\epsilon}_{\mathbf k}$) be identical to the
local lattice Green's function $G_{L}(\omega,T)$, i.e.
\begin{equation}
G_{c,\sigma}(\omega,T)=G_{L,\sigma}(\omega,T)\equiv
\sum_{\mathbf k}{1 \over \omega+\mu -\epsilon_{\mathbf k}
-\Sigma_{\sigma}(\omega,T)}
\ .
\label{eq:sc-dmft}
\end{equation} 
Here, $\Sigma_{\sigma}(\omega,T)$ is the conduction-electron self energy,
which is independent of ${\mathbf k}$ due to the local nature of the DMFT
approximation\cite{mh.89}.
In the following we drop spin indices for the Green's functions and the
self energy since, in this paper, we shall only be interested in
paramagnetic solutions of the Kondo lattice model.

To solve the effective single-impurity problem (\ref{eq:KM}) we use
Wilson's numerical renormalization group  (NRG)\cite{wilson.75+kww.80}
extended to finite-temperature spectra 
\cite{costi.92+94,hofstetter.00},
which is based on a logarithmic discretization that allows the low-energy
scales (such as the Kondo resonance) to be accessed with high resolution.
The self-consistency procedure involves: (i) guessing an initial
form for the effective medium $\tilde{\rho}(\omega)=
\sum_{\mathbf k}\delta(\omega-\tilde{\varepsilon}_{\mathbf k})$, (ii)
diagonalizing (\ref{eq:KM}) with the NRG and solving for its dynamics, 
(iii) re-calculating the effective medium by using Eq.~(\ref{eq:sc-dmft}). 
Steps (ii)-(iii) are repeated until self-consistency is reached.
Further details of the NRG and the DMFT procedure, which we refer 
to as DMFT(NRG), are given in \ref{app:details}
For an application of the DMFT(NRG) to the finite-temperature Mott
transition in the Hubbard model see Ref.~\onlinecite{bulla.01}.

In addition to the condition (\ref{eq:sc-dmft}) on the Green's
functions, we also fix the number of electrons $n_c$ in the conduction
band, by adjusting the chemical potential $\mu$ at each iteration, until
$n_c$ converges to the desired value.
This number is obtained in terms
of the local conduction electron density of states $\rho_{c}(\omega,T)$, as
\begin{equation}
n_{c} =
N_{\rm sites}^{-1}\sum_{j,\sigma} \langle c^\dagger_{j,\sigma} c_{j,\sigma}
\rangle = \int_{-\infty}^{+\infty}d\omega\;f_T(\omega)\rho_{c}(\omega,T)\;,
\label{eq:nc}
\end{equation}
where $f_T(\omega)=1/(1+\exp[(\omega-\mu)/T])$ is the Fermi function (the
Boltzmann constant and $\hbar$ are taken to be unity throughout the paper).
The interacting conduction electron density of states, in turn, is defined
by
\begin{equation}
\rho_c(\omega,T) = -{1 \over \pi} \; {\rm Im} \; G_{c}(\omega,T)\,.
\label{rhoc:eq}
\end{equation}
The chemical potential self-consistency is not implemented in
impurity-model calculations, but it is clearly required for 
concentrated systems, such as heavy fermions.

The quantity of main interest in this paper is the low-energy 
part of the $f$-electron spectral function 
of the Kondo lattice model $A(\omega,T)$. This is defined in 
\ref{app:spectra} It describes the Kondo resonance part
of the $f$-electron spectral function in the vicinity of the Fermi
level $E_{\rm F}$. The PES ($\omega < E_{\rm F})$ and IPES 
($\omega> E_{\rm F}$) intensities are given in terms of $A(\omega,T)$ by
multiplying this by the occupation factors for electrons $f_{T}(\omega)$ and
holes $1-f_{T}(\omega)$, respectively. Another useful quantity is defined 
in \ref{app:spectra}: $\chi_{zz}''(\omega,T)$, the spectral 
density of magnetic excitations for the effective quantum impurity model.

\subsection{The Role of the Band-Filling  $n_{c}$}
\label{filling}
We now discuss the significance of the other parameter
in the Kondo lattice model: the band filling $n_{c}$.
The $c_{{\mathbf k},\sigma}^{\dagger}$ fermions represent renormalized Bloch
states in the wide metallic bands of the crystal, with the correct symmetry
for coupling to the localized $f$ states.
Even in a stoichiometric compound, with integer number of electrons per unit
cell, the effective occupancy of the $c$ band need not be commensurate,
since the conduction electrons are shared between the $c$ band and other
bands of different symmetry, not included in the model.
Thus, $n_c=1$ represents the special case of particle-hole symmetry in this
effective band: it is appropriate for Kondo insulators (examples of which
include Ce$_{3}$Bi$_{4}$Pt$_{3}$, SmB$_6$, YbB$_{12}$, CeNiSn, and SmS
\cite{aeppli.92}), since $\rho_{c}$, $A$ and $\chi_{zz}''$ show true 
gaps at $T=0$.
$n_c\neq 1$ addresses naturally incommensurate filling of this effective
band.
In particular, $n_{c}\approx 1$ can be realized by doping the Kondo
insulators, e.g.\ by substitution of Ce with La in 
Ce$_{3}$Bi$_{4}$Pt$_{3}$\cite{hundley.94} or of Yb with Lu in 
YbB$_{12}$\cite{iga.85}.
Intermediate values of $n_c\approx 0.5$ describe metallic heavy fermions,
which we study in detail in Sect.~\ref{exp:sec}.
Finally, small values of $n_{c}$ should be relevant to
low carrier-density heavy-fermion systems, e.g.\ Yb$_{4}$As$_{3}$ 
and YbBiPt\cite{fisk.91}.

In addition to this direct significance of $n_{c}$ 
representing the actual number of electrons coupled to the $f$ states
in different materials, one can also attach a second meaning
to a variable
$n_{c}$: deviations of $n_c$ away from unity introduce particle-hole
asymmetry into the model. Such an asymmetry is generally present in the
Anderson lattice model and results from an asymmetric position of the $f$ 
level with respect to $E_{\rm F}$. In the Kondo lattice model, 
the $f$ level is projected out and real charge fluctuations are absent 
($n_{f}\equiv 1$). Although there is no $f$ level in the latter, one
can still partly simulate the effects of different $f$-level positions 
by changing instead the Fermi-level position $E_{\rm F}$, i.e.\
by varying $n_{c}$.
Within this scheme, Ce heavy-fermion compounds 
correspond to $n_{f}+n_{c}\leq 2$,
i.e.\ $0<n_{c}<1$, and Yb compounds are the particle-hole counterparts of
these with $1<n_{c}<2$.
Here, we only compute spectra in the range $0<n_{c}<1$, relevant for the
spectroscopy of Ce systems: the complementary region $1<n_{c}<2$, relevant
for Yb systems, is obtained by particle-hole symmetry exchanging
$\omega\rightarrow-\omega$ in the spectral functions.

%
%
%
\begin{figure}[t]
\centerline{\includegraphics[width=0.8\linewidth, height=0.6\linewidth]{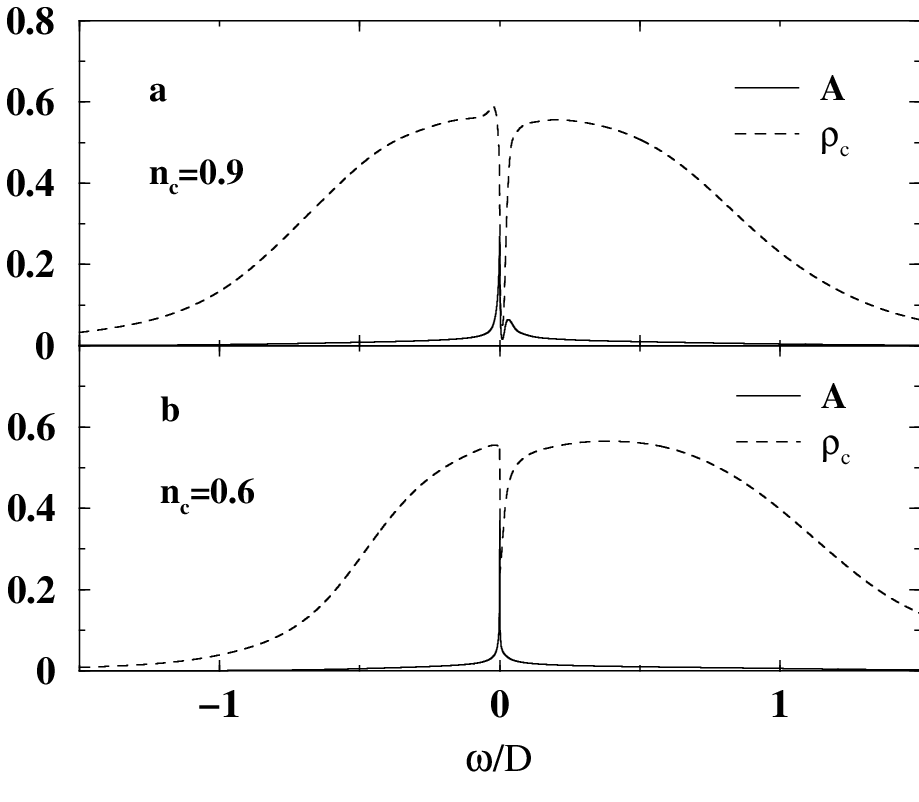}}
\protect\caption{
The spectral density $A(\omega,T=0)$ of the lattice Kondo resonance
(solid line) and of the conduction electrons $\rho_{c}(\omega,T=0)$ (dashed
line), for (a) $n_c=0.9$, and (b) $n_{c}=0.6$.
The detailed behavior near the Fermi level is shown in Fig.~\ref{fig2}.
\label{fig1}
}
\end{figure}

\section{RESULTS}
\label{results:sec}

All calculations are carried out for $J/D=0.3$, sufficiently smaller
than unity for the low-energy spectra to show universal scaling with the
Kondo-energy scale, but also sufficiently large to avoid unnecessarily
large numbers of NRG iterations in the region of small band filling, where
the Kondo scale becomes very small.
All spectra are measured with frequency $\omega$ relative to the chemical
potential $\mu$.

%
%
%
\begin{figure}[t]
\centerline{\includegraphics[width=0.8\linewidth, height=0.6\linewidth]{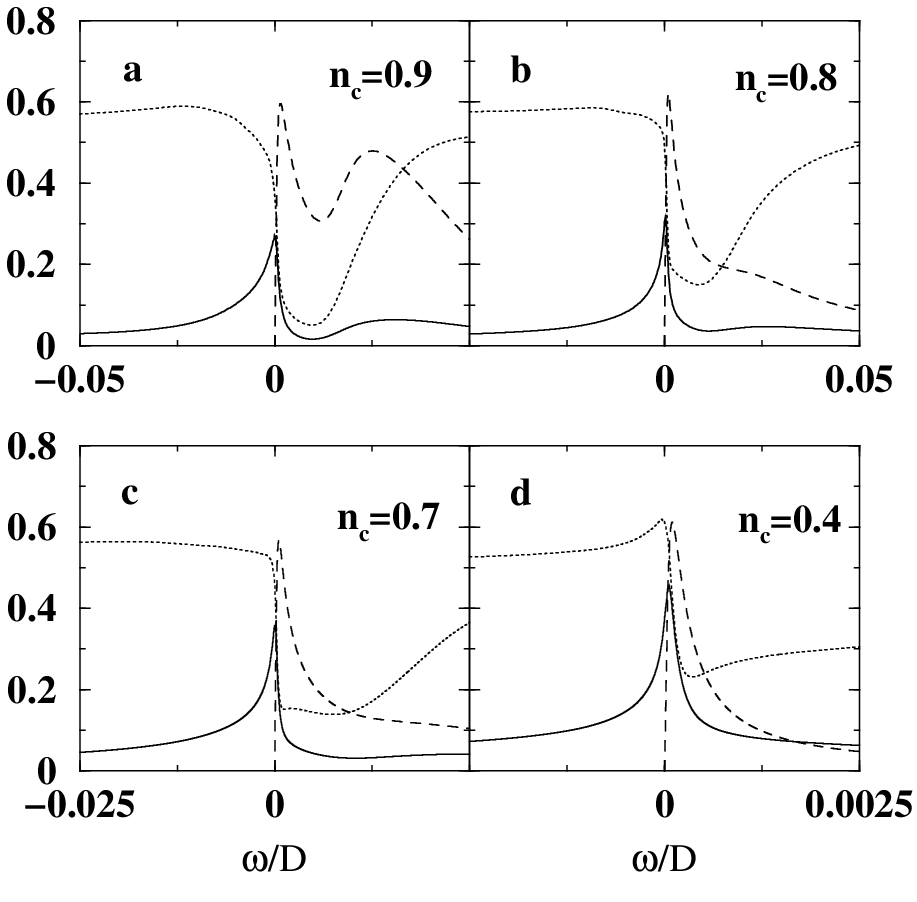}}
\protect\caption{
The $T=0$ conduction electron spectral density $\rho_{c}(\omega,0)$
(dotted line), the lattice Kondo resonance $A(\omega,0)$ (solid line), 
and the magnetic excitation spectrum $\chi_{zz}''(\omega,0)$
(dashed line), for decreasing band fillings: (a) $n_{c}=0.9$, (b)
$n_{c}=0.8$, (c) $n_{c}=0.7$, and (d) $n_{c}=0.4$.
The curves for $\chi''$ are rescaled arbitrarily for the purpose of
comparison with $\rho_{c}$ and $A$. Different frequency 
ranges are used in (a-d), as the lattice Kondo 
scale is decreasing with decreasing $n_{c}$ (see text).
\label{fig2}
}
\end{figure}

\subsection{ $T=0$ Spectral Densities and Low-Energy Scales}

Figure~\ref{fig1} shows the zero-temperature spectra for $\rho_c$ and $A$ on
a wide energy scale, of the order of the unperturbed bandwidth $W=2D$, and
for two fillings, $n_{c}=0.9$ and $n_{c}=0.6$.
The large-scale effects of interaction on $\rho_{c}$ are clearly visible:
band spreading, the opening of a renormalized hybridization pseudogap and
consequently a reduction in height of the conduction electron spectral
density relative to the unperturbed spectral density
(\ref{eq:semi-elliptic}).
These global effects, on the scale of the whole band depend on the
particular value of $J$ and $n_{c}$: we are interested instead in 
the low-energy behavior of the spectral functions, which, instead, 
is universal, i.e.\ independent of $J$ (but depend on $n_c$).

Figure \ref{fig2} zooms close to the Fermi level on the $T=0$ spectra,
plotting them for different band fillings $n_c$.
Two different energy scales emerge clearly in these spectra.

As soon as $n_{c}$ moves away from $1$ (Fig.~\ref{fig2}a), the $n_c=1$
insulating gap turns into a {\em pseudogap}, for all spectra.
We denote the characteristic width of this pseudogap in $\rho_{c}$ by
$T^{*}$. 
The pseudogap in $\rho_c$ lies just above the Fermi level for $n_{c}<1$
(and, by particle-hole symmetry, just below the Fermi level for $n_{c}>1$).
It persists for all $n_{c}$, but becomes increasingly asymmetrical and
step-like as $n_{c}\rightarrow 0$ (see Fig.\ \ref{fig2}b-d), making its 
precise value increasingly more difficult to specify accurately. 
Consequently, we have not made an exhaustive analysis of the 
dependence of $T^{*}$ on
$n_{c}$, but our calculations in the range 
$0.2\lesssim n_{c}\lesssim 0.96$ indicate that $T^{*}$ decreases 
slowly with decreasing $n_{c}$ and that it correlates approximately
with the single-ion Kondo scale, being a factor of $2-3$ larger than 
this, if we define the latter as the full width at half maximum (FWHM)
of the single-ion Kondo resonance at $T=0$. 
We find $T^{*}/D\approx 0.02 - 0.025$ for $n_{c}\gtrsim 0.6$. 
The pseudogap in $A$ manifests itself for $n_{c}$ close to 1.
The $A$-spectrum represents the lattice
Kondo resonance split by hybridization effects into a lower branch
pinned at the Fermi level and an upper branch at $\omega\approx T^{*}$.
The upper branch looses intensity at the expense of the main
branch at the Fermi level with decreasing $n_{c}$, and it
eventually disappears for $n_{c}\lesssim 0.7$.
Our conclusion is that in the $S=1/2$ Kondo lattice model, a
splitting of the Kondo resonance by hybridization effects is only
discernible close to half-filling ($n_{c}\gtrsim 0.7$), although
in principle such a splitting is present for all $n_{c}$ (see 
Ref.~\onlinecite{martin.82,kaga.88}).
The pseudogap is present also in the spectrum of the
magnetic excitations $\chi_{zz}''$: a peak at $\omega\approx T^{*}$
for $n_{c}$ close to half-filling represents excitations across a spin pseudogap.
On decreasing $n_{c}$, the peak in $\chi_{zz}''$ at $\omega \approx T^{*}$ 
becomes a shoulder at $n_{c}=0.8$, and then it disappears for 
$n_{c}\lesssim 0.7$, together with the upper branch of the lattice Kondo 
resonance in $A$.
In contrast, for $\rho_c$ the pseudogap persists for all $n_{c}$, so that 
$T^{*}$ remains a relevant scale for this quantity at all band fillings.

%
%
\begin{figure}[t]
\centerline{\includegraphics[width=0.8\linewidth, height=0.6\linewidth]
{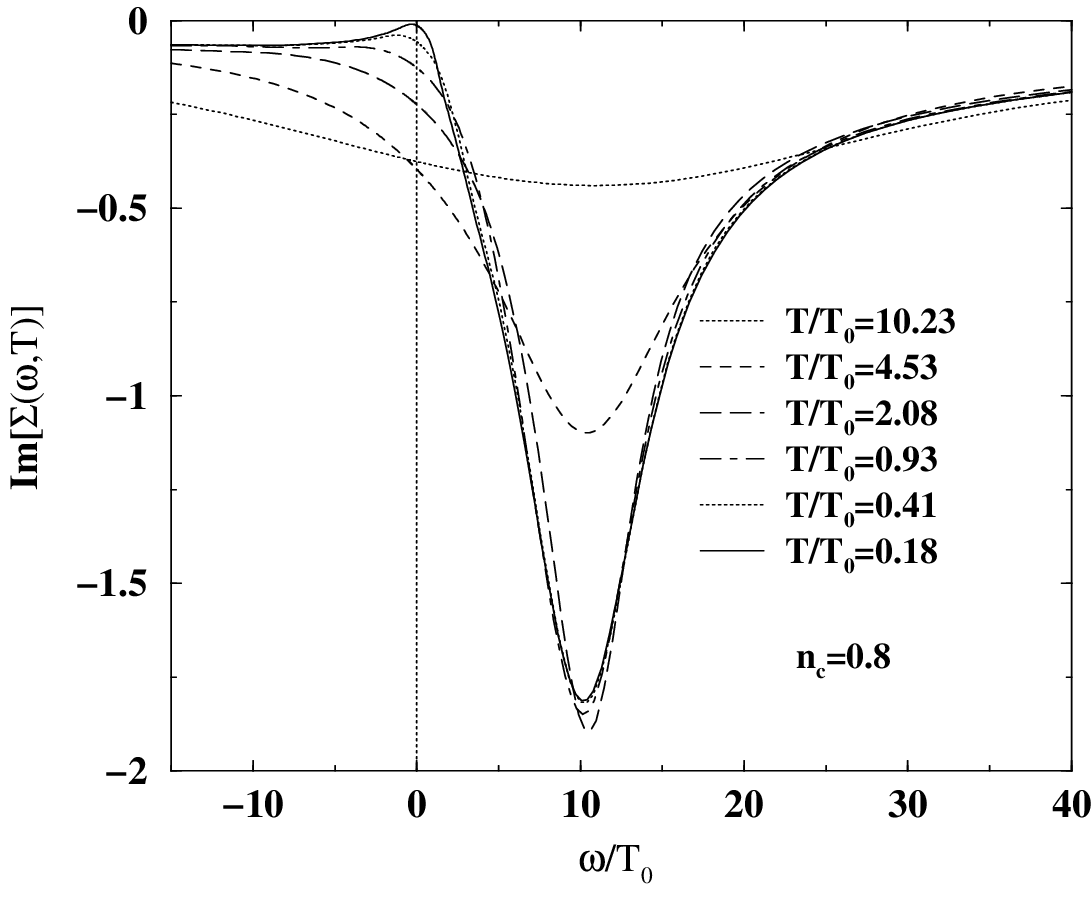}}
\protect
\caption{
Temperature dependence of the imaginary part of the self-energy 
Im$[\Sigma(\omega,T))]$ for $n_{c}=0.8$, showing that $T_{0}$ 
\protect\cite{note-T0values} is the Fermi liquid coherence scale.
\label{self-energy0.8}
}
\end{figure}

The second energy scale apparent in the spectra of 
Fig.~\ref{fig2} is associated with the onset of Fermi-liquid coherence 
(as will become clearer with the discussion of the self energy below).
It appears in Fig.~\ref{fig2} as the characteristic width of the $T=0$ 
lattice Kondo resonance in $A$ (i.e.\ its main branch at the Fermi level), 
the lowest peak in the magnetic excitation spectrum  $\chi_{zz}''$ and 
also as the energy range over which $\rho_{c}$ has a drop at the Fermi 
level. Clearly, there are several ways of defining this scale $T_{0}$.
In this paper we define it as the position of the
lowest-energy peak in the $T=0$ magnetic excitation spectrum $\chi_{zz}''$.
The Kondo scale $T_{0}$, defined in this way, has the same dependence on
$n_{c}$ and $J$ as that defined in other ways, but may differ from these by
factors of order unity (for example, the scale $T_K$ defined as one half of
the FWHM of the Kondo lattice resonance is 2-3
times larger than $T_0$).
We have compared the dependence of $T_{0}$ on $n_{c}$
\cite{note-T0values} with that of the single-ion Kondo scale, 
$T_{\rm imp}$ (defined in the same way as $T_{0}$, i.e., as the low-energy
peak position in the local dynamical susceptibility, and using the same
$J$ and $\rho_{0}$ as for the lattice model calculations). 
For $n_{c}\geq 0.4$, we find that $T_{0}<T_{\rm imp}$ and 
$T_{0}/T_{\rm imp}$ is consistent with the form $n_{c}\exp(p\,n_{c})$,
with $p\approx 5/2$, found
%
for the Anderson lattice model in the 
strong-correlation Kondo limit\cite{pruschke.00} within the 
DMFT(NRG) approach.
Burdin {\em et al.}\cite{burdin.00}, using a slave-boson mean field
approach, also find an increase of $T_{0}/T_{\rm imp}$ with $n_{c}$.

We find that, for fixed $n_{c}$, both $T_{0}$ and $T^{*}$ 
have the same $J$-dependent renormalization with decreasing $J$.
$T_{0}$ and $T^{*}$ are closest near half-filling and become increasingly
differentiated with decreasing $n_{c}$ (see also
Fig.~\ref{pseudogap0.8}-\ref{pseudogap0.4} below).

\subsection{$T$ Dependence of the Self-Energy}

Figure~\ref{self-energy0.8} shows the temperature and frequency dependence
of the imaginary part of the conduction electron self-energy
Im$[\Sigma(\omega,T)]$.
It vanishes at low temperatures at $\omega=E_{F}$ on a temperature 
scale of order $T_{0}$. 
This is consistent with the interpretation of $T_{0}$ as the Fermi-liquid
coherence scale.
The strong damping close to $\omega\sim T^*$ is associated with incoherent
excitations due to the pseudogap in $\rho_{c}$.
It correlates with the position of the broad high-energy peak in the
magnetic excitation spectrum  of Fig.~\ref{fig2}b.

%
%
\begin{figure}[t]
\centerline{\includegraphics[width=0.8\linewidth, height=0.6\linewidth]
{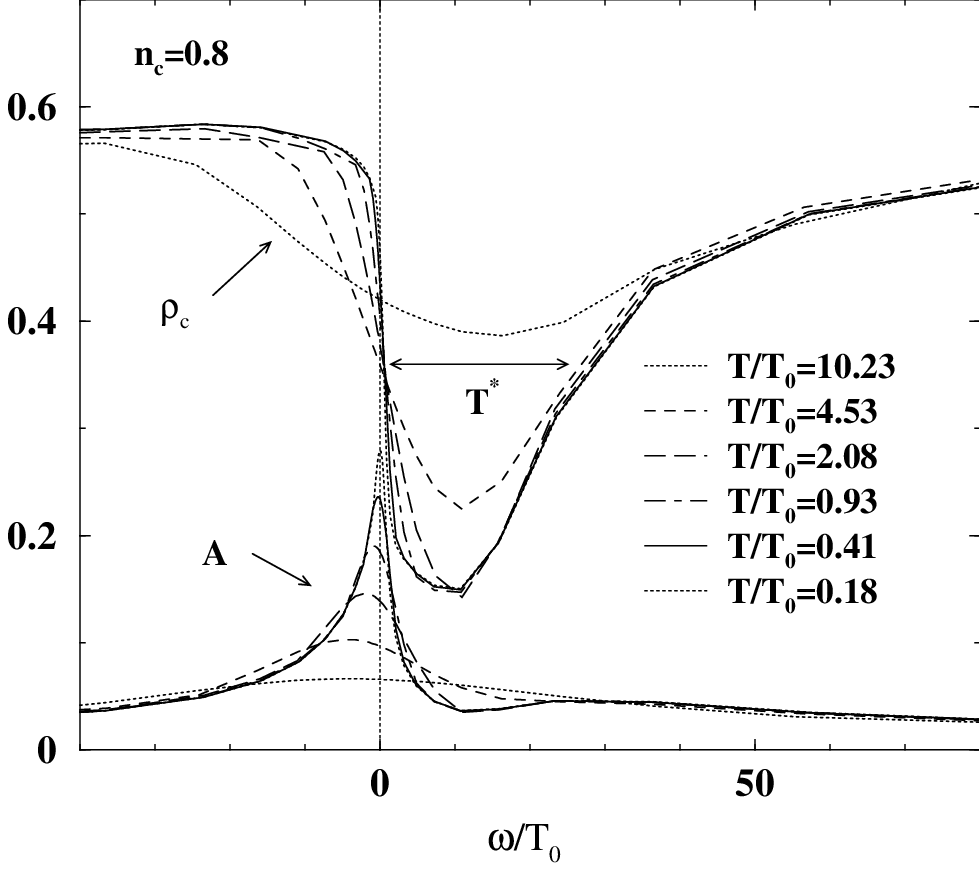}}
\protect
\caption{
Temperature dependence of $\rho_{c}$ for $n_{c}=0.8$ in the region near the
Fermi level and the pseudogap. The $A$-spectrum is shown for comparison.
Temperatures and energies are shown in units of 
$T_{0}$\protect\cite{note-T0values}.
\label{pseudogap0.8}
}
\end{figure}

%
%
\begin{figure}[t]
\centerline{\includegraphics[width=0.8\linewidth, height=0.6\linewidth]
{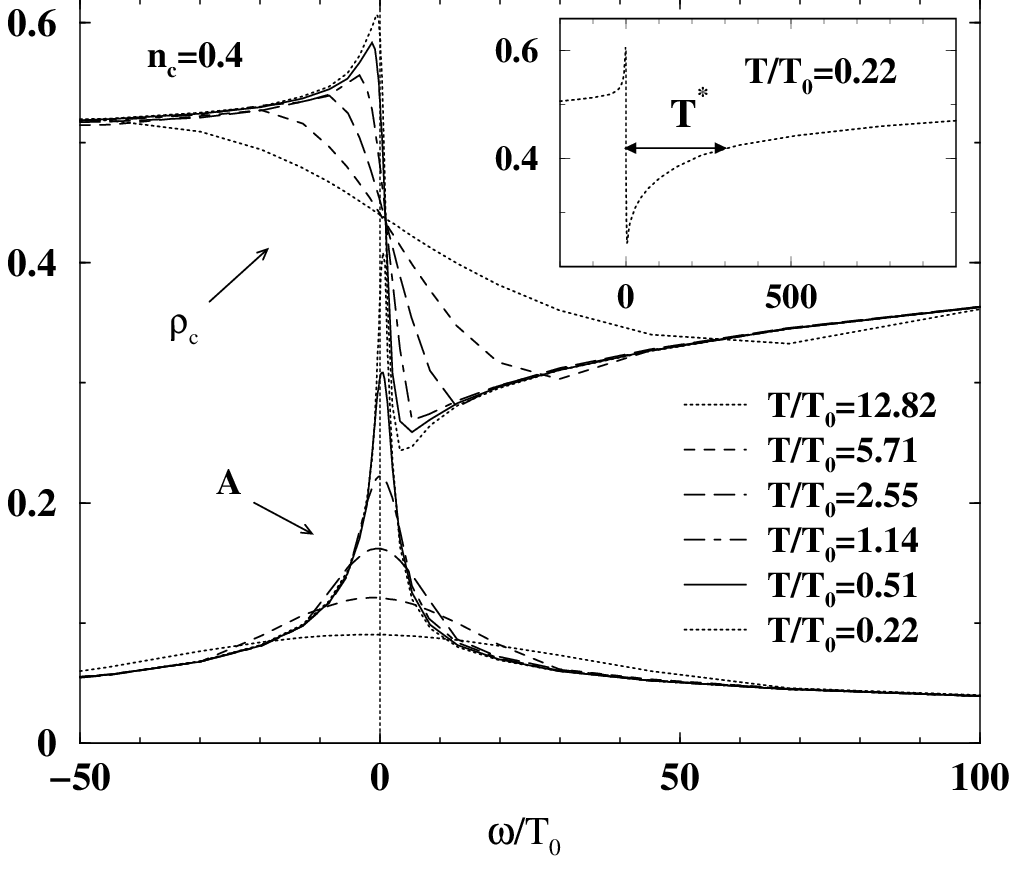}}
\protect
\caption{
Temperature dependence of $\rho_{c}$ for $n_{c}=0.4$ in the region near the
Fermi level and the pseudogap. The $A$-spectrum is shown for comparison. The
inset shows the pseudogap in $\rho_{c}$ on a larger scale. 
%
\label{pseudogap0.4}
}
\end{figure}

\subsection{$T$ Dependence of $\rho_{c}$ and $\chi_{zz}''$}

The role of the two energy scales $T^{*}$ and $T_{0}$ in the spectra
is further clarified in Fig.~\ref{pseudogap0.8}-\ref{pseudogap0.4}
which show the temperature dependence of $\rho_c$ for $n_{c}=0.8$
($T^{*}/T_{0}\approx 17$) and $n_{c}=0.4$
($T^{*}/T_{0}\approx 300$). 
The temperature dependence of $A$ is shown for comparison and
will be discussed in detail below.
We see that, as for the self-energy, the Fermi-liquid scale $T_{0}$ 
also sets the scale for the temperature dependence of $\rho_{c}$ close 
to the Fermi level (just as for $A$).
The feature of the $\rho_{c}$ spectrum directly affected on this
temperature scale is the sharp drop at $E_{\rm F}$.
Instead, the region near the minimum of the pseudogap in $\rho_{c}$ 
(and, whenever it exists, also in $A$ and $\chi_{zz}''$) starts 
to evolve at temperatures larger than $T_0$.
It finally closes at higher temperatures $T\gtrsim T^{*}$.

\begin{figure}[t]
\centerline{\includegraphics[width=0.8\linewidth, height=0.6\linewidth]
{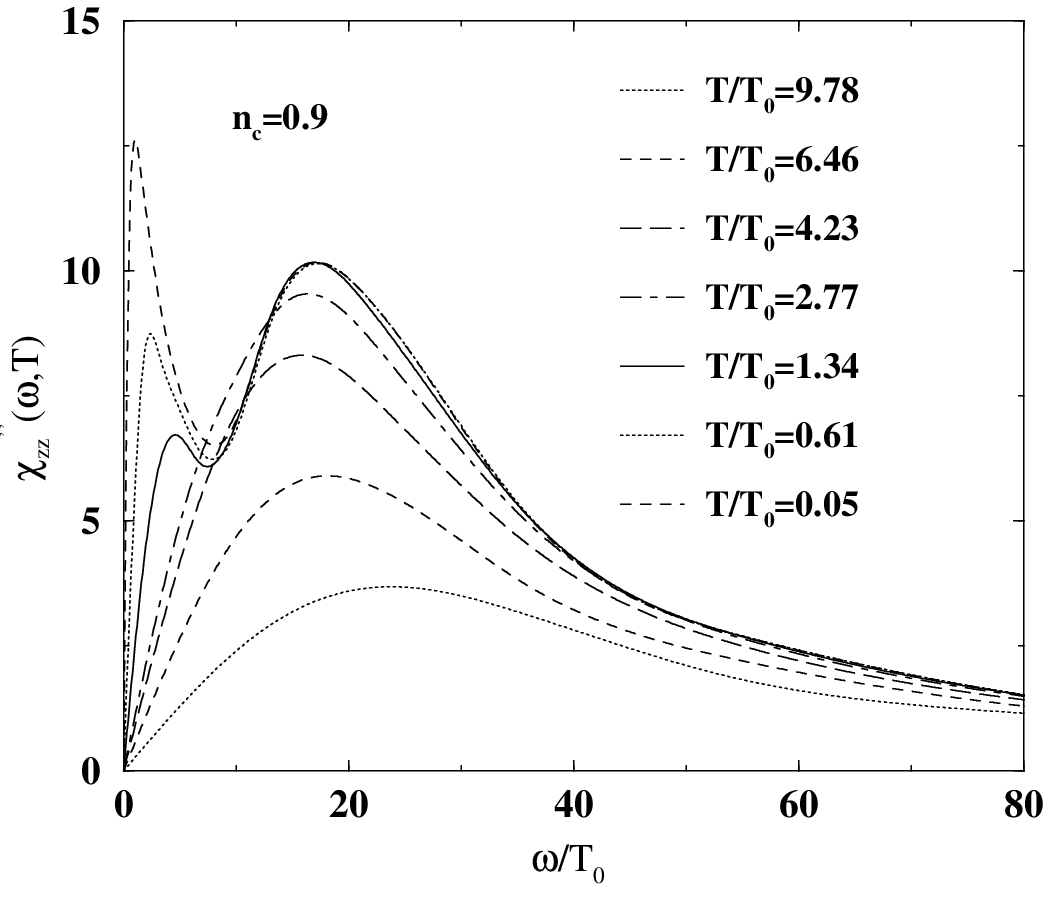}}
\protect
\caption{
Temperature dependence of $\chi_{zz}''$ for $n_{c}=0.9$. This type
of response is similar to the Kondo insulating behavior measured for
CeNiSn \protect\cite{mason.92} and 
Ce$_3$Bi$_4$Pt$_3$\protect\cite{severing.91}.
%
\label{chi0.9}
}
\end{figure}

For band fillings $n_{c}\gtrsim 0.7$, the pseudogap is also evident in
$\chi_{zz}''$. Figure~\ref{chi0.9} shows that $T_{0}$ sets the
scale for the temperature dependence of a quasielastic peak, whereas
excitations across a spin pseudogap (of order $T^{*}$) give rise
to a peak in $\chi_{zz}''$ at $\omega \sim T^{*}$.
The temperature dependence of the latter is set by the scale $T^{*}$. This
type of response also characterizes the Kondo insulating behavior at
$n_{c}=1$, as long as $T>0$, so that the gap behaves effectively as a
pseudogap at finite temperature, as seems typical of correlated
insulators\cite{jarrell.95}.
The neutron scattering results for CeNiSn\cite{mason.92} and
Ce$_3$Bi$_4$Pt$_3$\cite{severing.91} show qualitatively the same
temperature dependence.

%
%

%
%
\begin{figure}[t]
\centerline{\includegraphics[width=0.8\linewidth, height=0.6\linewidth]
{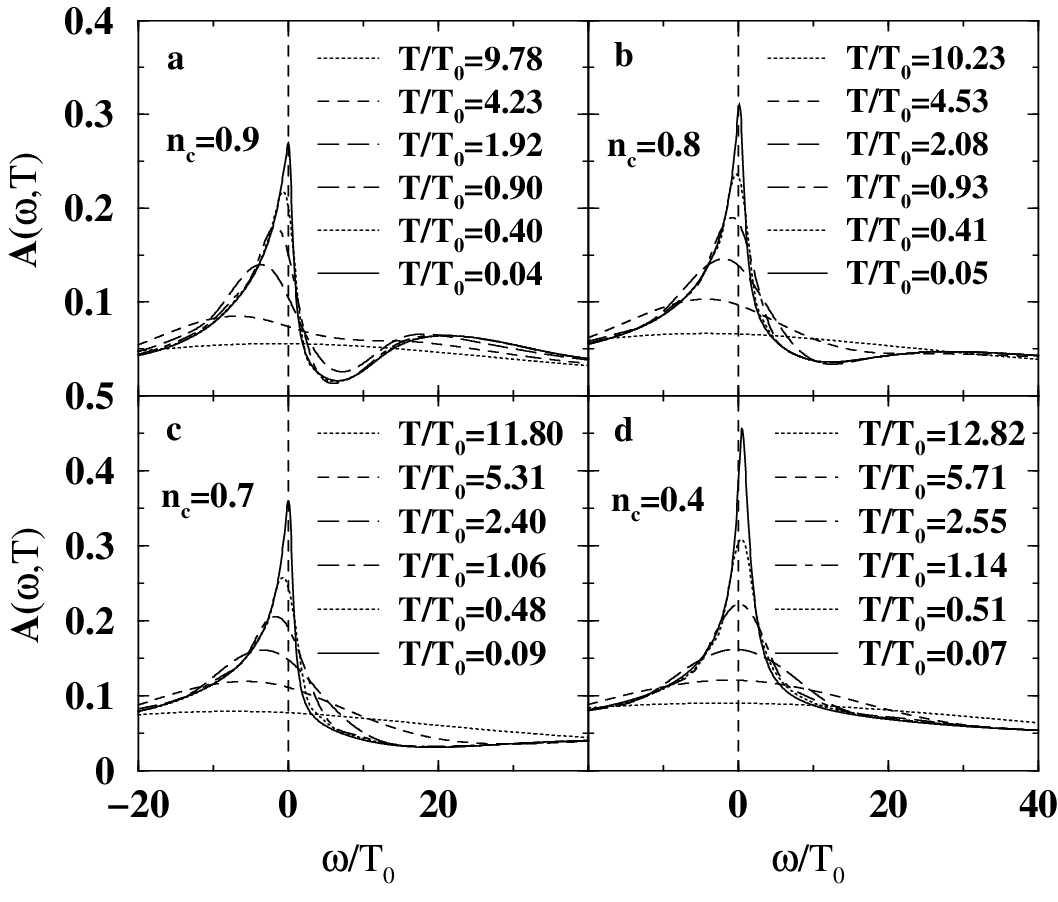}}
\protect
\caption{
Temperature dependence of the lattice Kondo resonance 
for the same band fillings as in \protect{Fig.~\ref{fig2}}.
%
%
Temperatures and energies are shown in units of $T_{0}$, where $T_{0}$ 
is the lattice Kondo scale appropriate for each $n_{c}$ and defined 
in the text \protect\cite{note-T0values}.
\label{fig3}
}
\end{figure}

%
%
%
%
\begin{figure}[t]
\centerline{\includegraphics[width=0.8\linewidth, height=0.6\linewidth]{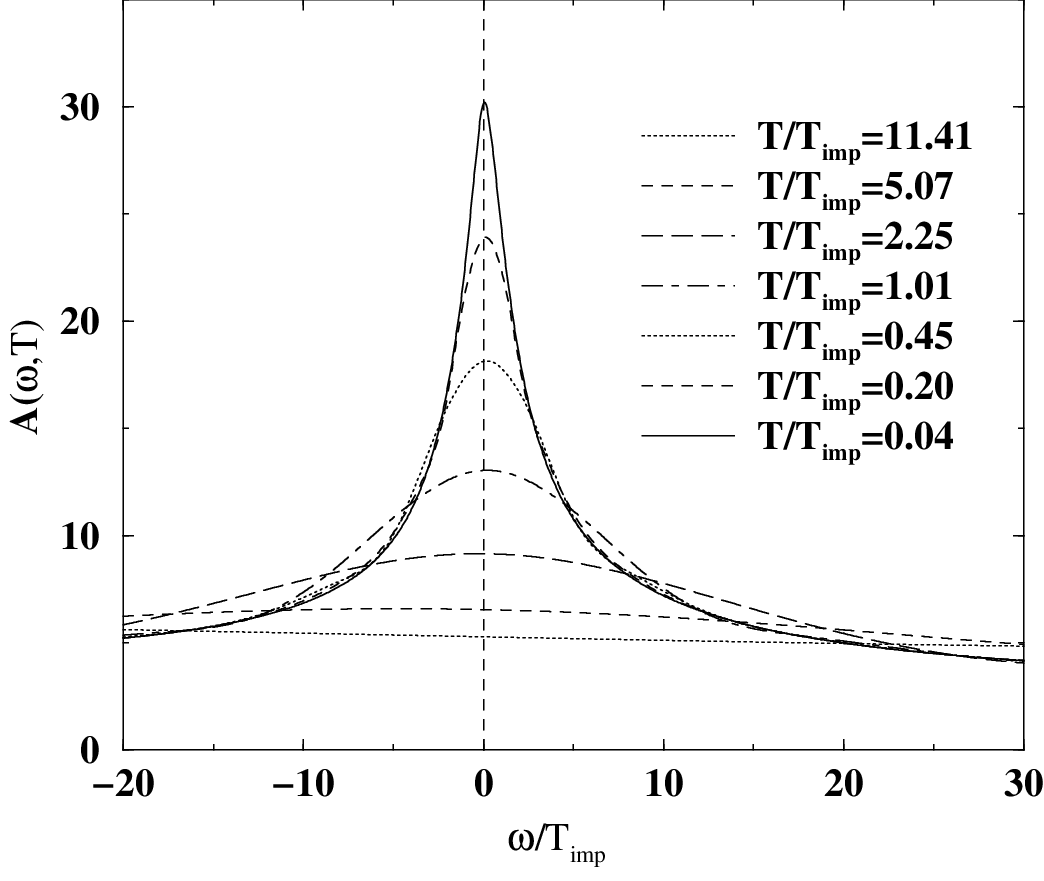}}
\caption{
Temperature dependence of the impurity Kondo resonance calculated from the
AIM in the Kondo regime by using the NRG\protect\cite{costi.92+94}.
The local level position used is 
$\varepsilon_{d}=-4\Delta$ and the Coulomb interaction $U=4\pi\Delta$,
with $\Delta=0.1D$ the hybridization strength and $D=1$ the half-bandwidth
for a flat conduction band $\rho_{0}(\omega)=1/2D$.
The Kondo scale, 
$T_{\rm imp}=4.22\times 10^{-4}D$, is defined for the impurity model in 
the same way as for the lattice model, as the peak position 
in the $T=0$ magnetic excitation spectrum.
\label{fig-am-kr}
}
\end{figure}

\subsection{Discussion of the Two Energy Scales}

At high temperatures, approaching the scale $T^{*}$, the spectra 
are similar to those of a Kondo impurity, namely the split Kondo
resonance in $A$ for $n_{c}\geq 0.7$ reverts to a single-ion 
like Kondo resonance at the Fermi level whose height decreases with 
increasing temperature (see Fig.~\ref{fig3}a), $\chi_{zz}''$ has a
peak at $\omega\approx T^{*}$ whose height also decreases with
increasing temperature, and $\rho_{c}$ is approximately independent of
temperature. The physics appears like that of a collection of 
quasi-independent Kondo scatterers, since $T^{*}$ correlates 
with the single-ion Kondo scale describing the high-temperature 
properties of these Kondo scatterers.
On decreasing the temperature, the effects of lattice
coherence become quite pronounced for $T\approx T^{*}$: the spectra 
develop a pseudogap resulting from the formation of hybridized bands 
between the $f$ and conduction states. 
Fermi-liquid coherence only sets in
at much lower temperatures, specifically for $T\ll T_{0}$.

The formation of the paramagnetic heavy-fermion state
therefore appears to proceed in two stages, (i), a single-ion Kondo
effect at $T\gg T^{*}$ in which the $f$-electrons are not strongly
hybridized with the itinerant conduction electron states, so that they
play little role in screening, and, (ii), a low-temperature Kondo effect at 
$T<T^{*}$ in which the itinerant states have
a strong $f$-electron character (via formation of renormalized 
hybridized bands) so that the latter also contribute to screening of
the local moments. This second stage Kondo effect leads to the 
reduced lattice Kondo scale $T_{0}$ describing just the width 
of the main branch of the split Kondo resonance
at the Fermi level. Two energy scales, a single-ion and lattice Fermi-liquid 
Kondo scale, have also been found within a recent slave-boson treatment 
of the Kondo lattice model\cite{burdin.00}.

\subsection{$T$ Dependence of the Kondo Resonance}

Figure~\ref{fig3} shows the temperature dependence of the lattice
Kondo resonance for several band fillings.
Compare the behavior of the main branch of the lattice Kondo 
resonance away from half filling with that calculated within the AIM 
\cite{costi.92+94}, and shown in Fig.~\ref{fig-am-kr}.
The lattice Kondo resonance, like the impurity one, is pinned
to just above the Fermi level for $T\ll T_{0}$. However, in contrast to the
impurity case, the peak position $E_{m}(T)$ relative to the chemical
potential {\em shifts to below} the Fermi level on a scale $\lesssim T_{0}$
with increasing temperature.
This effect is particularly evident slightly away from half-filling
(e.g. $n_{c}=0.9$ in Fig.~\ref{fig3}a), and it becomes less pronounced as
the band filling decreases toward zero, but is still significant as
compared to the impurity-model predictions (Fig.~\ref{fig-am-kr}).
For the band fillings shown in Fig.~\ref{fig3}, $n_{c}\geq 0.4$, the
lattice Kondo resonance vanishes on a scale of approximately $10\;T_{0}$.
This is the same scale over which the single-impurity Kondo resonance
vanishes (see Fig.~\ref{fig-am-kr}). 
We therefore do not see a significantly slower dependence (termed
``protracted behavior''\cite{tahvildar-zadeh.97})
of the height of the lattice Kondo resonance as compared to the impurity
one.
This contradicts the $n_{c}=0.4$ PES calculations for the 
periodic Anderson model of Ref~\cite{tahvildar-zadeh.98}.
Our calculations cannot rule out, however, that such ``protracted''
behavior might be present in other quantities, such as in the static
susceptibility, or that such behavior will appear in spectral densities
for $n_{c}\ll 1$ (the ``exhaustion'' limit\cite{nozieres-exhaustion}, an
important conceptual limit that shall require clarification:
we have not addressed this here, because we are mainly interested in
metallic heavy fermions with appreciable band fillings).
%
For a recent
experimental evaluation of the ``protracted screening'' scenario see
Ref.~\onlinecite{lawrence.01}.

%
%
%
\begin{figure}[t]
\centerline{\includegraphics[width=0.8\linewidth, height=0.6\linewidth]
{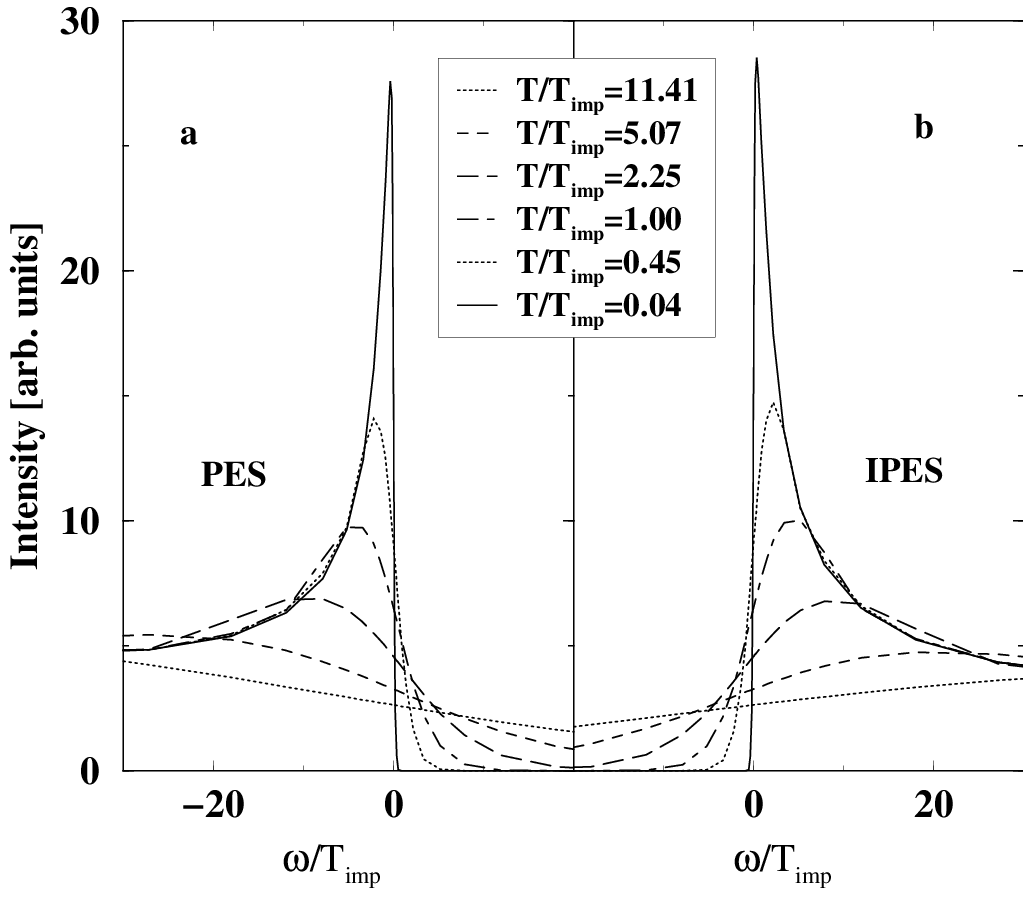}}
\protect\caption{
Temperature dependence of the low-energy PES intensity (left) and IPES
intensity (right) for the impurity Kondo resonance calculated from the AIM
in the Kondo regime using the NRG \protect\cite{costi.92+94}. 
The same temperatures as in Fig.~\ref{fig-am-kr} are used.  
\label{fig-am-int}
}
\end{figure}

%
%
%
%
\begin{figure}[t]
\centerline{\includegraphics[width=0.8\linewidth, height=0.6\linewidth]
{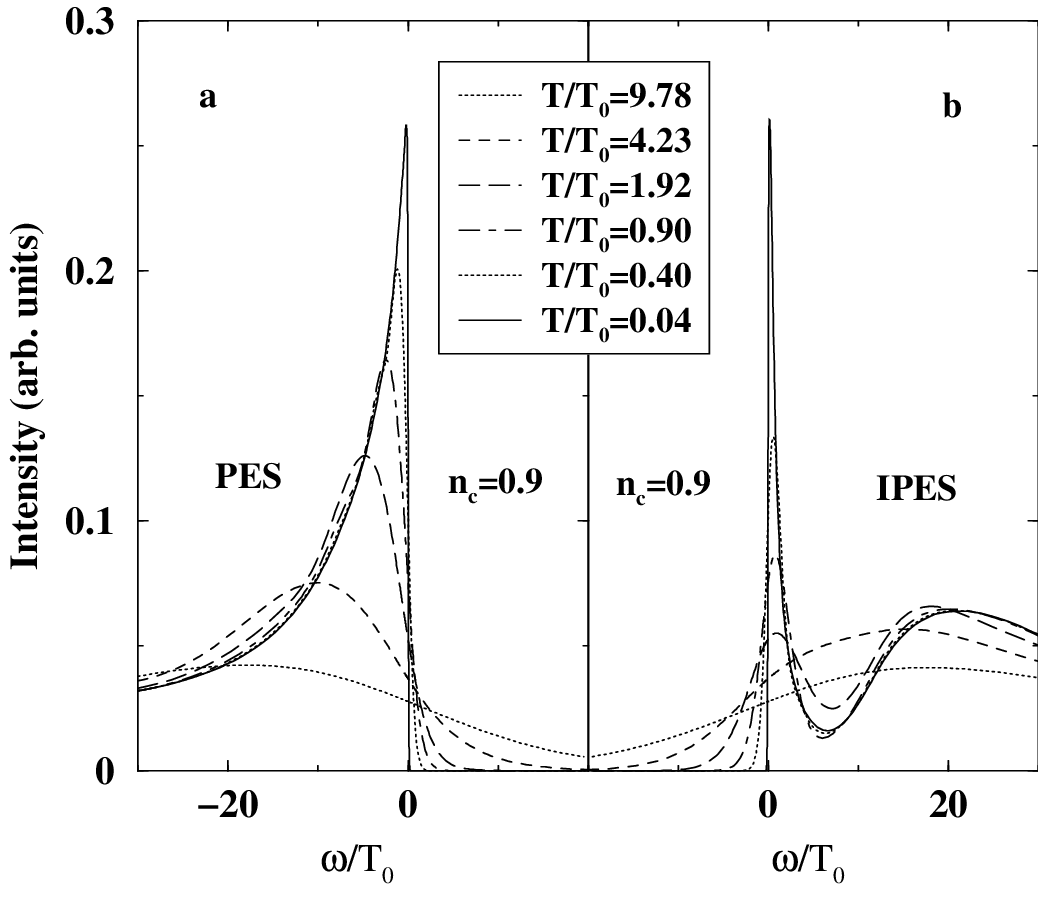}}
\protect\caption{
Temperature dependence of the low-energy PES intensity (left) and IPES
intensity (right) for the lattice Kondo resonance for $n_{c}=0.9$ (rather
close to the Kondo insulating case $n_{c}=1$). The temperatures shown
correspond to those in Fig.~\ref{fig3}a.
\label{fig-int-0.9}
}
\end{figure}

%
%
%
\begin{figure}[t]
\centerline{\includegraphics[width=0.8\linewidth, height=0.6\linewidth]
{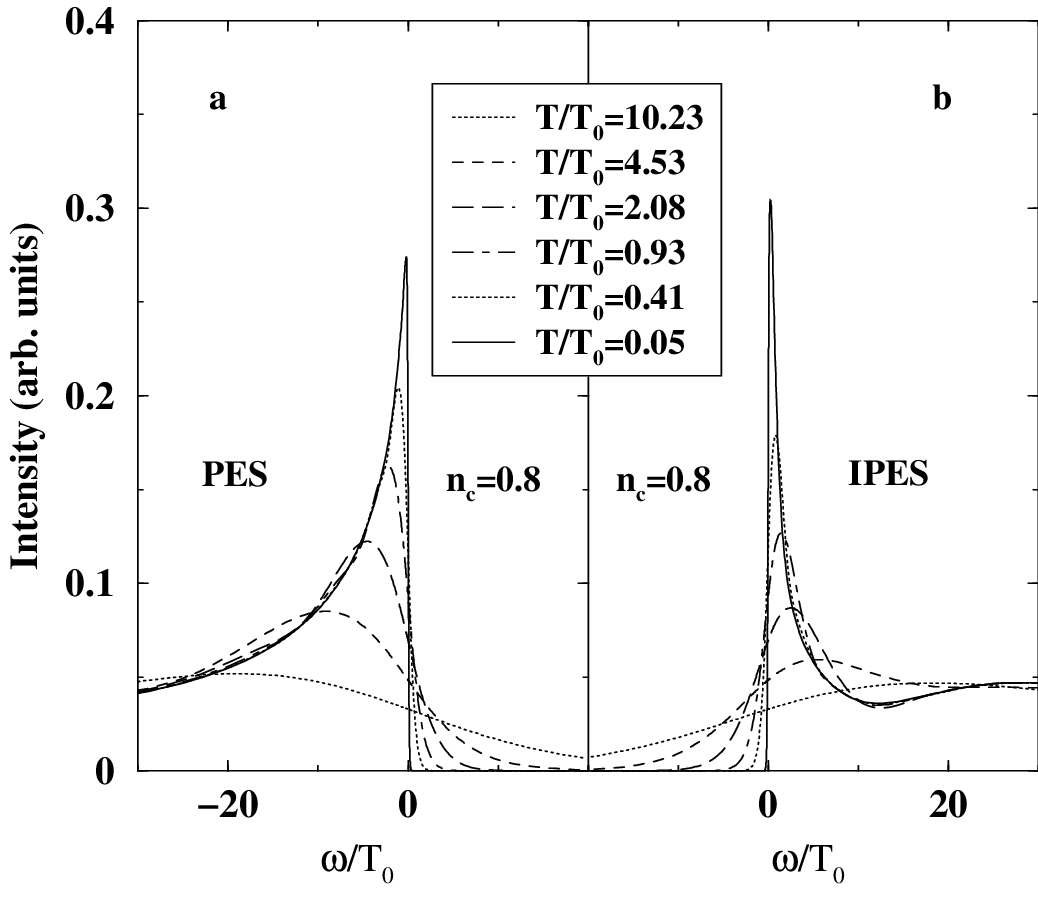}}
\protect\caption{
Temperature dependence of the low-energy PES intensity (left) and IPES
intensity (right) for the lattice Kondo resonance for $n_{c}=0.8$. The
temperatures are those of Fig.~\ref{fig3}b.  
\label{fig-int-0.8}
}
\end{figure}

%
%
%
%

%
%
%
%
\begin{figure}[t]
\centerline{\includegraphics[width=0.8\linewidth, height=0.6\linewidth]
{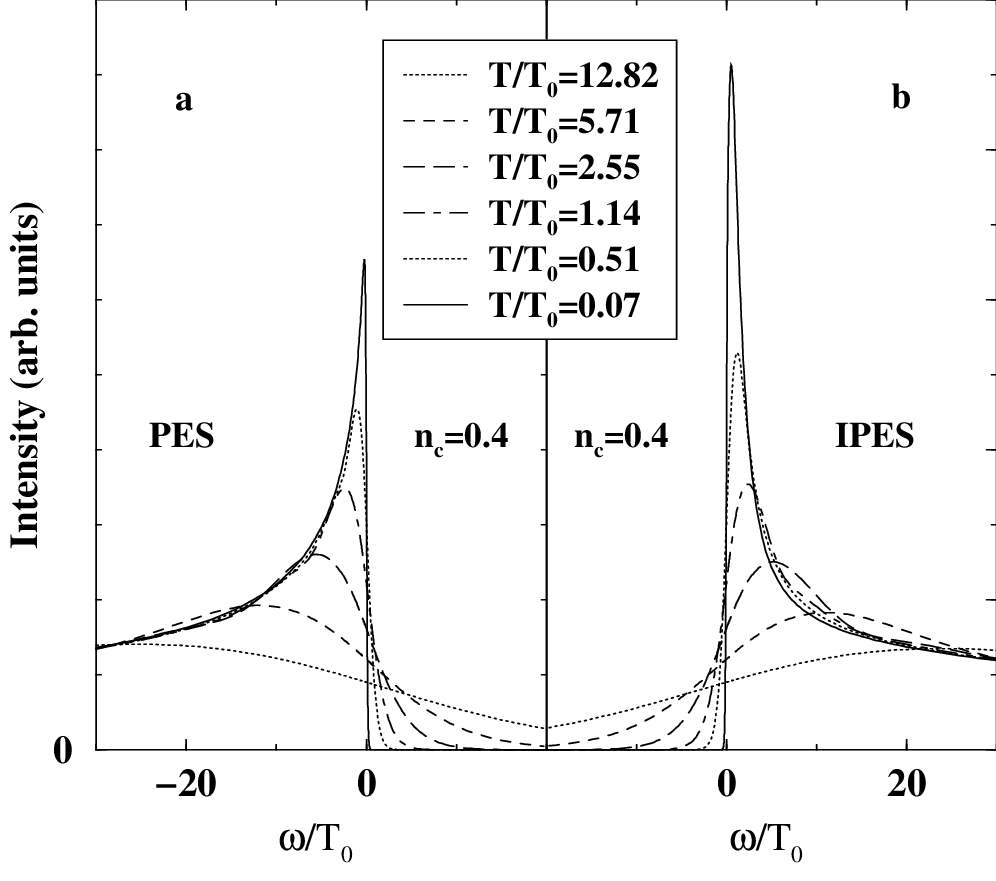}}
\protect\caption{
Temperature dependence of the low-energy PES intensity (left) and IPES
intensity (right), for the lattice Kondo resonance for $n_{c}=0.4$. The
temperatures are those of Fig.~\ref{fig3}d.
\label{fig-int-0.4}
}
\end{figure}

\section{$T$ DEPENDENCE OF PES AND IPES}
\label{int:sec}

The low-energy PES ($I_{-}$) and IPES ($I_{+}$) intensities
are defined by
\begin{eqnarray}
I_{-}(\omega,T) &=& f_T(\omega)A(\omega,T)\label{eq:pes}\\
I_{+}(\omega,T) &=& (1-f_T(\omega))A(\omega,T).
\label{eq:ipes}
\end{eqnarray}
For comparison with the Kondo lattice model predictions, we first outline
the PES and IPES spectra from the AIM.

\subsection{$T$ Dependence of PES and IPES for the AIM}

A typical $T$-dependent Kondo resonance from the AIM is shown in
Fig.~\ref{fig-am-kr}.
The derived PES and IPES are shown in Fig.~\ref{fig-am-int}.
It is often stated that the PES intensity in Ce compounds is very
small compared to the IPES intensity because the Kondo resonance has
most of its weight above the Fermi level.
For the present $N=2$ AIM we
see that the Kondo resonance is, in fact, pinned very close to the
Fermi level. Correspondingly, we observe nearly equal intensities in
PES and IPES spectra in Fig.~\ref{fig-am-int}.
This does not contradict many
PES and IPES experiments on Ce compounds, however, since most of these
experiments are carried out at temperatures and resolutions far above 
the typical energies of the crystal field excitations or Kondo scale. 
As a result, they probe the full $N=6$ degeneracy
of the Ce $4f^{1}$ configuration and therefore cannot be described within 
an AIM with $N=2$.
Including the orbital degeneracy in the AIM is known to shift the position
of the Kondo resonance to further above the Fermi level\cite{hewson.93},
and therefore to increase the asymmetry between the PES and IPES
intensities, as observed in many Ce (and Yb) heavy fermions.
Nevertheless, in systems where the $N=2$ AIM approach is applicable, 
the spectra of
Fig.~\ref{fig-am-int} suggest that experiments at sufficiently high
resolution and low enough temperatures (such that only the low-energy
doublet is effectively playing a role) should give nearly equal
intensities in PES and IPES for a system with a doublet ground state.

\subsection{$T$ Dependence of PES and IPES for the Kondo Lattice Model}

The PES and IPES intensities for the Kondo lattice model are more
complicated and show a range of behaviors, depending on band filling $n_c$.
Fig.~\ref{fig-int-0.9}-\ref{fig-int-0.4} shows the PES and IPES intensities
for the four band fillings $n_{c}=0.9$, $n_{c}=0.8$, $n_{c}=0.7$ and 
$n_{c}=0.4$ of Fig.~\ref{fig3}.
For Kondo insulators ($n_{c}=1$), there is experimental evidence on 
Ce$_3$Bi$_4$Pt$_3$\cite{breuer.98} and CeRhAs\cite{kumigashira.01}
indicating the development of a gap in the PES on decreasing 
temperature. This is consistent with our predictions for the $A$-spectrum 
at $n_{c}=1$, to be published elsewhere, and also with 
calculations on the symmetric Anderson lattice model 
\cite{jarrell.95,rozenberg.96}.

The results for $n_{c}$ close to 1 (Fig.~\ref{fig-int-0.9}) are
relevant to doped Kondo insulators, e.g. to hole-doped Ce systems,
such as Ce$_{3-x}$La$_{1-x}$Bi$_{4}$Pt$_{3}$\cite{hundley.94}.
As noted earlier (see Fig.~\ref{fig3}), the Kondo resonance is highly
asymmetric for $n_{c}\lesssim 1$ with most of its weight lying below the
Fermi level.
Consequently, the bulk of the intensity in Fig.~\ref{fig-int-0.9} lies
in the PES part of the spectrum. In addition the strong shift of the
Kondo resonance from above to below the Fermi level on increasing
temperature for the case $n_{c}\lesssim 1$, 
noted in Sect.~\ref{results:sec},
further diminishes the near-$E_{F}$ intensity in IPES as compared to 
that in the PES.

Another interesting feature of our calculations is that the pseudogap can
be seen as a dip in the IPES intensity at $\omega \approx T^{*}$, which
becomes increasingly well developed with decreasing temperature.
Our predictions for hole-doped Ce systems apply equally well, by
particle-hole symmetry, to electron-doped Yb systems.
Results for the electron-doped Kondo insulator 
Yb$_{1-x}$Lu$_{x}$B$_{12}$\cite{susaki.97} indeed indicate
an increase in the photoemission intensity at the Fermi 
level upon doping YbB$_{12}$ with Lu.

On further decreasing the band filling
(Fig.~\ref{fig-int-0.8}-\ref{fig-int-0.4}),
the IPES intensity increases with respect to the PES intensity. This
results from the more symmetric lattice Kondo resonance together with the
tendency of this resonance to move further above the Fermi level
for smaller $n_{c}$. In addition, the pseudogap in the IPES intensity 
fills in below $n_{c}\approx 0.7$.
For $n_{c}=0.4$, far enough from 
the Kondo insulating state to be representative of metallic heavy fermion 
systems, we find a behavior rather similar to the AIM spectra, i.e.\ the 
IPES intensity is slightly larger than the PES intensity at low 
temperatures, $T\ll T_{0}$.
However, a small increase of the temperature to $T\gtrsim T_{0}$ makes these
intensities nearly equal (e.g.\ Fig.~\ref{fig-int-0.4}).
Inclusion of the orbital degeneracy of the $f$-level should lead
to a displacement of the lattice Kondo resonance further above the Fermi
level, as occurs in impurity models, and consequently to larger IPES
intensities for Ce systems than PES intensities.

\section{COMPARISON WITH EXPERIMENTS}
\label{exp:sec}

Ideally, our predictions for the $S=1/2$ Kondo lattice model are
suitable for heavy-fermion systems with a local doublet ground state
characterized by a separation $\Delta_{\rm CF}$ from the first
crystal-field excitation much larger than the Kondo scale $T_{0}$.
An example of such a system would be CeCu$_{2}$Si$_{2}$, 
which has $T_{0}\sim 6$~K and $\Delta_{\rm CF}\sim 30$~meV 
\cite{goremychkin.93}. 
Photoemission spectra have been measured for this compound\cite{reinert.01b}
at $6$~meV resolution (FWHM). The absolute resolution is high, but the 
relative resolution for probing the Kondo resonance is still quite 
low due to the small Kondo scale. Moreover, most measurements were 
taken at temperatures $T\gtrsim 10 T_{0}$, with only one temperature
($T=15$~K) relevant for testing predictions about the Kondo resonance.
In this respect, the experiments on YbAgCu$_{4}$\cite{weibel.93} and 
YbInCu$_{4}$\cite{moore.00} are more promising, since measurements were taken 
at $T$ both above and below the relevant Kondo scales. 
Unfortunately, for these systems, neutron scattering studies
\cite{lawrence.97,severing.90} indicate that the 
crystal-field excitation energy is comparable to the Kondo scale
($\Delta_{\rm CF}\sim T_{0}$).
In the temperature range of the experiments $T\sim 10-200$~K, thermal 
population of the crystal-field states will be important and 
would require further refinements of our $S=1/2$ model, which, however,
are beyond the scope of the present work. 
With this proviso, we compare our results, which
include lattice coherence and correlation effects, with measurements 
of the temperature dependence of the low-energy $f$-spectrum for 
YbInCu$_4$\cite{moore.00} and YbAgCu$_4$\cite{weibel.93}. 

YbInCu$_4$ exhibits an isostructural phase transition, similar to that of
Ce, with a 0.5\% volume change and a $0.1$ change in $f$-level
occupancy at a temperature $T_{v}=42$~K.
Both phases show the characteristic features of paramagnetic heavy-fermion
behavior.
The high-temperature phase has a Kondo scale $T_{0}=30$~K, whereas the 
low-temperature phase has a much larger Kondo scale $T_{0}=400$~K
\cite{sarrao.96}.
Temperature-dependent photoemission measurements were carried out in both
phases\cite{moore.00} ($T=$12--40~K for the low-temperature phase, 
and $T=$50--150~K for the high-temperature phase).

YbAgCu$_4$ shows a single paramagnetic heavy-fermion phase down to the
lowest temperatures\cite{rossel.87}.
Neutron-scattering experiments indicate a quartet crystal-field
ground state with
a Kondo scale $T_{0}=100$~K \cite{severing.90}, consistent with 
thermodynamic measurements\cite{besnus.90}. 
%
Measurements for this compound were carried out in the temperature 
range 20--260~K and show a strong temperature dependence 
of the near-$E_{\rm F}$ feature in qualitative agreement with expectations 
from Kondo phenomenology\cite{weibel.93}.

The experiments on ${\rm YbInCu_{4}}$ used single crystals in
contrast to the polycrystals used for ${\rm YbAgCu_{4}}$.
In both experiments, one observes a near-$E_{\rm F}$ feature (due to
excitations from the $f^{7/2}$ ground state of the ${\rm Yb}$ ions),
interpreted as the Kondo resonance, and an $f^{5/2}$ spin-orbit side band
at approximately 1~eV below $E_{\rm F}$.
The bulk and surface contributions to these excitations can be separated in
the experiments, allowing bulk effects to be investigated.
The $f^{5/2}$ spin-orbit side-band is not included in our model, but its
absence will not affect the overall predictions for the $f^{7/2}$ excitation 
in a range of $\ll 1$~eV around the $E_{\rm F}$ spectrum.

The experimental resolution ($\Delta E=25$~meV
FWHM) is included by convoluting the calculated
intensities (described in Sect.~\ref{int:sec}) with the
appropriate Gaussian broadening.
Since the resolution is comparable to or below typical Debye
energies, some effect of phonon broadening on the spectra may be
observable\cite{joyce.92,moore.00}.
Specifically, for both ${\rm YbInCu_{4}}$ and ${\rm YbAgCu_{4}}$, we take a
Gaussian phonon broadening term of the form $\Delta E_{ph}=\alpha T$
reaching 50~meV at 150~K \cite{moore.00}: the total width used
to convolute the calculated intensities is then given by 
\begin{equation}
\Delta E_{tot}=\left(\Delta E_{ph}^{2}+\Delta E^{2}\right)^{1/2} \ .
\label{DeltaEtot}
\end{equation}

For all the comparisons, we use an intermediate value $n_c = 0.4$.
This choice is motivated by the requirement of being in a regime
representative of metallic heavy fermions, characterized by an asymmetric
position of the localized $f$ states.
%

%
%
%
%
\begin{figure}[t]
\centerline{\includegraphics[width=0.8\linewidth, height=0.6\linewidth]{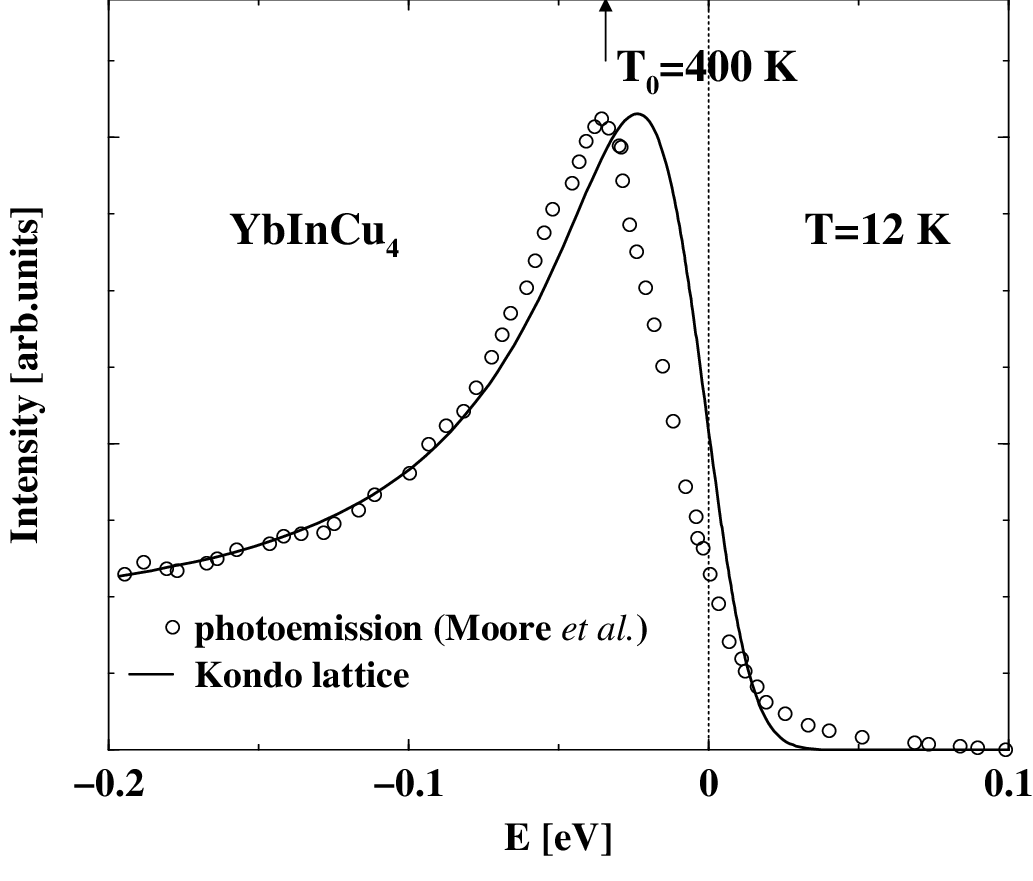}}
\protect\caption{
Comparison of the calculated PES intensity for the lattice Kondo resonance 
with the experimentally measured intensity \protect\cite{moore.00}
 for the $f^{7/2}$ near-$E_{\rm F}$ 
feature in ${\rm YbInCu_{4}}$ at $T=12K\ll T_{0} = 400$~K. 
The calculated curve includes the $\Delta E=25$~meV Gaussian 
instrumental resolution.
}
\label{fig-YbInCu4.12}
\end{figure}

\subsection{Low-Temperature Phase of YbInCu$_4$}

In Fig.~\ref{fig-YbInCu4.12} we compare the calculated
and experimental intensities for YbInCu$_4$ at the lowest
temperature $T=12$~K at which measurements were taken in the
low-temperature phase (the curves are normalized so that the peak heights
coincide).
Since the Kondo scale $T_{0}=400$~K is much larger than the temperature,
this measurement essentially probes the $T=0$ lattice Kondo
resonance below the Fermi level, convoluted with a Gaussian 
of width $\Delta E=25$~meV.
All other measurements in this low-temperature phase are for temperatures
$T<T_{v}=42$~K $\ll T_{0}=400$~K and can also be regarded as essentially
zero-temperature measurements.
%
At such low temperatures, the Kondo resonance is fully developed, so no
temperature dependence is expected, and indeed almost none is seen
\cite{moore.00}.

Figure~\ref{fig-YbInCu4.12} shows reasonable agreement 
between the calculations and the experimental results.
In particular, the theory reproduces the correct width of the 
near-$E_{\rm F}$ feature, as well as the slow decay of the 
intensity at negative energies.
We attribute the latter to the non-Lorenzian decay of the Kondo 
resonance at high energies similar to that found for the impurity
Kondo resonance\cite{frota.86} 
(whose lineshape for $|\omega|\gg T_{0}$ is of the Doniach-Sunjic type).
The position of the peak in the intensity at $T\ll T_{0}$ is related to the
position $E_{m}$ of the Kondo resonance in $A(\omega,T)$.
 From Fig.~\ref{fig3} we estimate that 
$E_{m}(T\rightarrow 0)/T_{0}\approx -0.5$ for
the present calculations at $n_{c}=0.4$.
The experiments would indicate a value closer to $-1$, i.e.\ larger in 
magnitude than the $S=1/2$ Kondo lattice model result.
The predictions of the AIM in the Kondo regime (with occupancy $\approx 1$)
would be in far worse agreement, since from Fig.~\ref{fig-am-kr}
we see that the Kondo resonance is pinned even more closely to the
Fermi level than in the Kondo lattice calculations.
It is important to note, however, that there is no fundamental 
relationship between $E_{m}$ and the Kondo scale. $E_{m}$
simply reflects a particle-hole asymmetry in the model, but on physical
grounds one expects it to lie within approximately $T_{0}$ of 
the chemical potential. Its precise value, both for impurity and 
lattice models, depends on specific details influencing the particle-hole
asymmetry, such as the ground-state degeneracy of the $f$ state, 
the presence of low-lying crystal-field states, and the amount
of mixed valence as reflected in the $f$-level occupancy (fixed 
to $n_{f}=1$ in the Kondo lattice model)\cite{note-nhole}.
As our model describes a local ground-state doublet with $n_{f}=1$, it 
would be surprising if it could capture the precise value of $E_{m}/T_{0}$ for 
YbInCu$_{4}$, which would depend on such details. Hence the position 
of the peak in the intensity at $T\ll T_{0}$ is also not 
captured precisely.
It is likely that a Kondo lattice model, including the $f$-orbital 
degeneracy, or the corresponding Anderson lattice model, will 
show complete agreement between theory and experiment.

%
%
%
%
\begin{figure}[t]
\centerline{\includegraphics[width=0.8\linewidth, height=0.6\linewidth]
{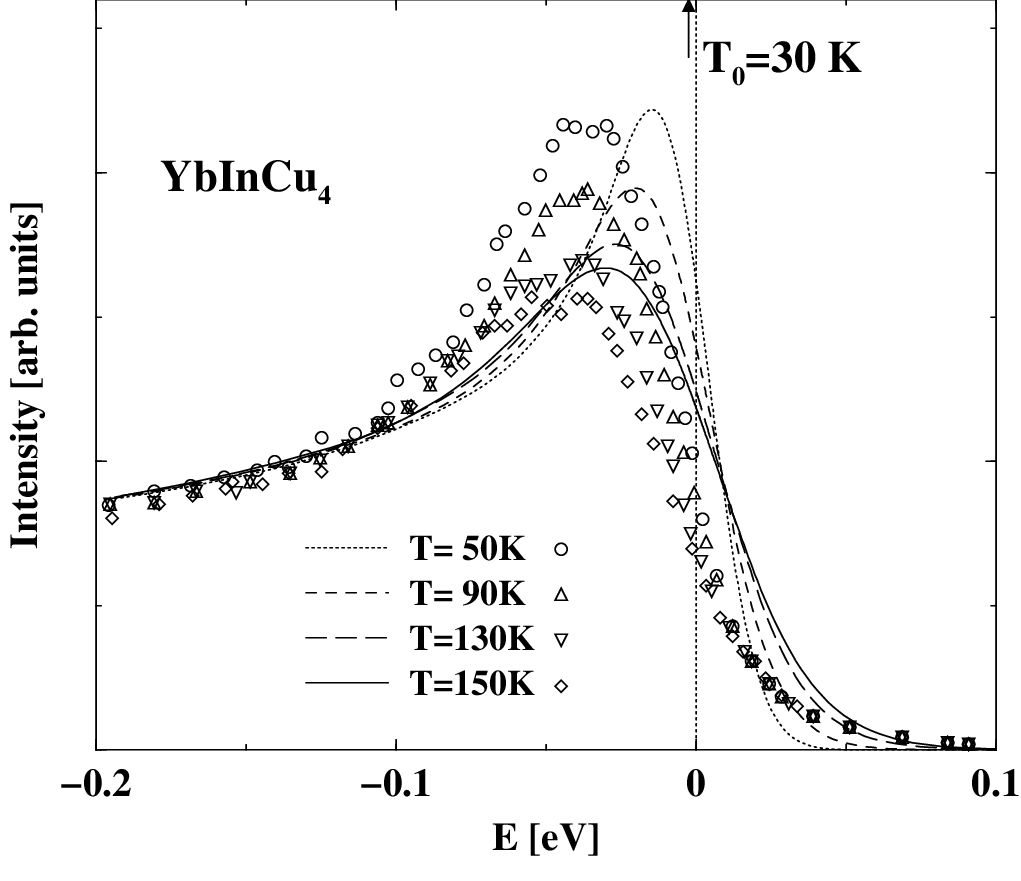}}
\protect\caption{
Comparison of the temperature dependence of the calculated PES intensity 
for the lattice Kondo resonance (lines) with the experimentally measured 
intensity (symbols) for ${\rm YbInCu_{4}}$ \protect\cite{moore.00}. 
The measurements were for the high-temperature 
phase, with a Kondo scale of $T_{0}=30$~K.
We compare only with the
$f^{7/2}$ near-$E_{\rm F}$ feature (the $f^{5/2}$ spin-orbit side band at 
$-1.375{\rm eV}$ in the experiments is not included in the present model).
The calculated curves are Gaussian convoluted to account for the instrumental
and phonon broadening $\Delta E_{tot}$ of Eq.~(\ref{DeltaEtot}).  }
\label{fig-YbInCu4-hiT}
\end{figure}

\subsection{High-Temperature Phase of YbInCu$_4$}

We turn now to the high-temperature phase of YbInCu$_{4}$, 
for which $T_{0}=30$~K and measurements have been taken up to
150~K$=5\;T_{0}$ \cite{moore.00}. This range of temperatures is 
interesting, because the results of the previous section predict
a strong temperature dependence in the near-$E_{\rm F}$ spectra 
in the range $0\leq T \leq 10\,T_{0}$ due to the Kondo effect.
A comparison of the measured intensities with the Kondo lattice 
model predictions is shown in Fig.~\ref{fig-YbInCu4-hiT}. 
As in Fig.~\ref{fig-YbInCu4.12}, the
resolution and Kondo scales are taken from experiment and no attempt
was made to obtain a best fit to the spectra by adjusting $T_{0}$
or $\Delta E$.
A phonon broadening reaching 50~meV at 150~K is included, as described
above.
This improves agreement for the two highest temperatures, but phonon
broadening is not the main source of temperature dependence in the
spectrum,  which is instead the temperature dependence of the Kondo
resonance in the spectral function $A(\omega,T)$.
The experimental spectra are consistent with this interpretation, because
they show a loss in intensity of the near-$E_{\rm F}$ feature as the
temperature is increased from 50~K to 150~K:
this agrees with the loss in spectral weight of the Kondo resonance with
increasing temperature, and cannot be attributed to phonon broadening,
which conserves the total spectral weight.

The lineshape is well reproduced by the calculations, as is the temperature
dependence of the intensity.
The curves are normalized so that the lowest-temperature peak heights in
theory and experiment agree.
This normalization gives agreement between theory and experiment for the
tails and peak heights at all temperatures.
Note that, at finite temperatures, the position of the intensity peak 
no longer reflects the Kondo resonance position (and therefore
also not the Kondo scale). Thermal broadening of the spectral 
function $A(\omega,T)$ in $I_{-}$, will shift the 
intensity peak to lower energies with increasing temperature, 
as observed in the calculations. The remaining discrepancy 
is essentially due to the oversimplified 
description of the $f$-states, resulting in too small a value 
for $E_{m}(T)$, as discussed above for the low temperature case. 

%
%
%
\begin{figure}[t]
\centerline{\includegraphics[width=0.8\linewidth, height=0.6\linewidth]{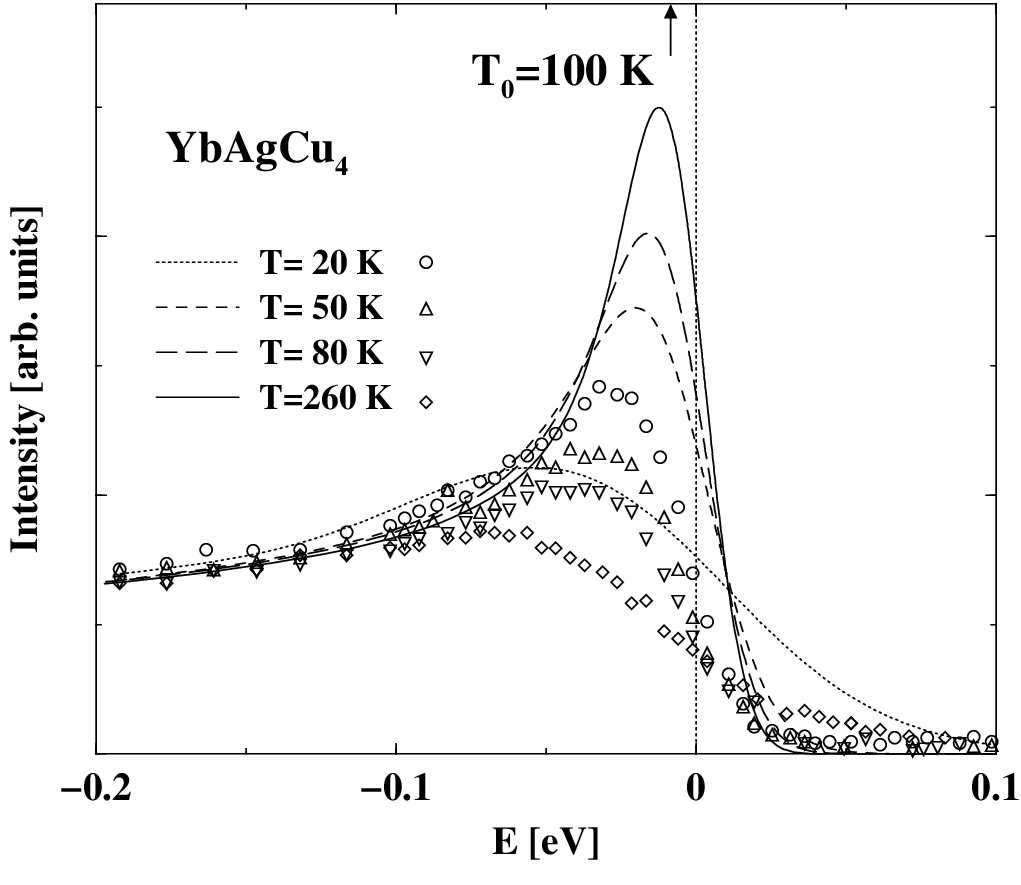}}
\protect\caption{
Temperature dependence of the low-energy PES intensity of 
YbAgCu$_4$ \protect\cite{weibel.93} compared to Kondo lattice calculations. 
Phonon broadening and instrumental resolution are accounted for as for
YbInCu$_4$ in Fig.~\ref{fig-YbInCu4-hiT} }
\label{fig-YbAgCu4}
\end{figure}

\subsection{YbAgCu$_4$}

Figure~\ref{fig-YbAgCu4} shows a comparison of the present
calculations to early experimental results of Weibel {\it et al.}
\cite{weibel.93} for the temperature dependence of the
near-$E_{\rm F}$ feature in YbAgCu$_4$.
As for YbInCu$_4$, we take the experimental resolution and Kondo scale
without any adjustment.
The agreement is less good than for YbInCu$_4$, but still
consistent with the interpretation of the temperature dependence of the
spectrum as arising mainly from the Kondo effect.
The discrepancy in the peak position
has the same origin as that described above for YbInCu$_4$,
and it could be overcome by a more realistic modeling of the $f$ states
of this compound.

For this compound, we normalize the curves to the tails in the intensity at
negative energies.
That part of the spectrum has much weaker temperature dependence, it
reflects the non-Lorenzian tail of the Kondo resonance and it is the
safest to use in case there are extrinsic (non-Kondo) temperature
dependences close to $E_{\rm F}$.
In contrast to the case of YbInCu$_4$, we do not obtain agreement for
both the tails and the peak height of the intensity as a function of
temperature.
We conclude that these early data, although indicating a Kondo effect
in the temperature dependence of the spectrum close to the Fermi level,
are not fully consistent with our model.
To investigate the origin of the discrepancy between theory and experiment,
it would be interesting to repeat such experiments, particularly on single
crystals as for YbInCu$_4$\cite{moore.00}.

\section{CONCLUSIONS}
\label{conclusions:sec}

In summary, we have studied the temperature dependence of the
single-particle spectral densities of the Kondo lattice model and their
evolution with band filling within dynamical mean field theory.
We found that the spectra exhibit two energy scales: a 
renormalized hybridization
pseudogap scale $T^{*}$, which evolves from the true hybridization gap of
the Kondo insulating state at $n_{c}=1$ on moving away from half-filling,
and a typically much smaller Fermi-liquid coherence scale $T_{0}$, 
characterizing the formation of the main branch of the split lattice 
Kondo resonance at the Fermi level. 
The weight of the upper branch of this lattice Kondo resonance at 
$\omega\approx T^{*}$ becomes negligible for $n_{c}\lesssim 0.7$.
The scale $T^{*}$ was found to correlate with 
the single-ion Kondo scale. 
The high-temperature behavior of the local spectral densities 
is similar to that expected for a collection of 
independent Kondo impurities. 
We found that the lattice Kondo resonance develops over the same
temperature range, $0\leq T \lesssim 10T_{0} < T^{*}$, as in 
the single-impurity models, at least for the range of band fillings
studied in this paper: $0.4 \lesssim n_{c}\lesssim 0.9$.
In this respect, the Kondo effect in this low-temperature phase, 
$T<T^{*}$, appears similar to the single-ion Kondo effect. 
However, the electrons involved have a strong $f$
character arising from the renormalized bands formed below 
$T\approx T^{*}$. 
We interpret the above findings as indicating a two-stage Kondo
effect for the formation of the paramagnetic heavy-fermion state:
a single-ion Kondo effect at $T\gg T^{*}$, and a low-temperature
Kondo effect at $T<T^{*}$ resulting from the modified electronic
structure below $T^{*}$, and giving rise to the new scale $T_{0}$.
Lattice coherence, associated with pseudogap
formation on a scale $T^{*}$, sets in at temperatures 
well above the Fermi liquid coherence scale $T_{0}$.

The results for the temperature dependence of the lattice Kondo 
resonance were used to calculate the low-energy part of the $f$-level 
PES and IPES intensities of heavy-fermion systems.
We made detailed predictions for the spectra of doped Kondo insulators
($n_{c}\lesssim 1$) and metallic heavy fermions. For the former, we 
found that most of the intensity of the Kondo resonance lies in the 
PES, and not in the IPES. 
This applies to hole-doped Ce or electron-doped Yb Kondo insulators. 
The predicted pseudogap in the $f$-electron spectrum is best seen in
the PES of electron-doped Yb Kondo insulators.
For hole-doped
Ce systems, this would lie just above the Fermi level in the IPES spectrum.
On moving away from the Kondo insulating toward the 
metallic heavy-fermion case, we found that the main intensity of the 
Kondo resonance shifts to the IPES spectrum and the PES and IPES
intensities appear qualitatively similar to impurity-model calculations. 

We made a parameter-free comparison of our results with high-resolution
PES data for YbInCu$_{4}$ and YbAgCu$_{4}$, taking
the Kondo scales and resolution from experiment.
Good agreement was found for the lineshape and temperature dependence of
YbInCu$_{4}$, although we argued that a more realistic treatment of the
$f$-states (including, for example, orbital degeneracy) would be 
required to obtain the precise peak position in the
PES intensity of the near-$E_{\rm F}$ feature.
In the experimental spectra we found evidence of some phonon broadening at
higher temperatures, but the main temperature dependence of the
near-$E_{\rm F}$ spectrum is due to the Kondo effect.

The present study has focused attention on the paramagnetic
metallic state of the $S=1/2$ Kondo lattice model and has
revealed some of the physics of this state, such as the relevant
energy scales involved and how these appear in the PES and IPES
spectra of heavy fermions. 
It would be interesting to identify these scales in other properties, 
such as in the optical conductivity, transport properties and in
the thermodynamics. 
Work on this is in progress and will be reported elsewhere. 
\section*{ACKNOWLEDGMENTS}

Peter W\"{o}lfle has played a key role in encouraging work by one
of the authors (TAC) on correlated electron systems, and it is a 
great pleasure to dedicate this paper to him on the occasion of his 60th 
birthday.
We thank the Newton Institute - Cambridge and the ILL - Grenoble for support
during the development of this project.
NM also acknowledges support from the SISSA-Trieste and Italian INFM.
We wish to thank Y. Baer, N. Brookes, D. Moore and S. J. Oh for 
useful discussions.

\appendix
\section{NRG AND DMFT DETAILS}
\label{app:details}

We provide, here, some of the details of the NRG and the DMFT applied 
to the Kondo lattice model.

Starting from Eq.(\ref{eq:KM}) we follow the standard procedure for
casting the effective impurity model into a linear chain form suitable 
for an NRG calculation\cite{wilson.75+kww.80}.
This first involves
a logarithmic discretization of the conduction band about the chemical
potential $\mu(T)$ (which for the purposes of this Appendix is taken as the
zero of energy).
The discretized energies are $\pm\;
D_{\pm}\;\Lambda^{-n},\;n=0,1,\dots$, 
where $\pm D_{\pm}$ are the upper ($+$)
and lower ($-$) band edges, respectively, and $\Lambda>1$ is the
discretization parameter, set to $1.5$ in the calculations.
A unitary transformation of these states then yields a new tridiagonal set 
$b_{n,\sigma},n=0,1,\dots$, such that $b_{0,\sigma}=c_{0,\sigma}$.
The Hamiltonian takes the following form:
\begin{eqnarray}
&&{\cal H}_{imp} = J\;\sum_{\mu,\nu}{\mathbf S}\cdot b_{0,\mu}^{\dagger}
{{\bm \sigma}}_{\mu,\nu} b_{0,\nu}\nonumber\\ 
&&+ {\sum_{n=0,\nu}^{\infty}\varepsilon_{n} b_{n,\nu}^{\dagger} b_{n,\nu}
+ \sum_{n=0,\nu}^{\infty}\lambda_{n} (b_{n,\nu}^{\dagger} b_{n+1,\nu}
+H.c.)}.
\label{eq:KM-chain} \
\end{eqnarray}
We use this form for the actual calculations, which involve iteratively
diagonalizing Eq.~(\ref{eq:KM-chain}) in the standard NRG manner 
\cite{wilson.75+kww.80}, i.e.\ adding the orbitals $n=0,1,\dots$ successively
and retaining only the lowest energy eigenstates at each stage.
We retained $462$ non-degenerate states at each NRG iteration and we 
checked that the spectra did not change significantly by further increasing
the number of kept states.
The spectra are calculated at the natural NRG temperature scales,
$T_{N}\sim \lambda_{N}$\cite{note-temps}.

This NRG technique is the basic ingredient in the solution of the Kondo
impurity problem at the core of the each DMFT step, then iterated to
approach self consistency.
We terminate the DMFT iteration when Eq.~(\ref{eq:sc-dmft}) is satisfied
for $\rho_c$ to an accuracy of $10^{-4}$ in the relative norm of successive
iterates (i.e.\ $||\rho_c^n-\rho_c^{n-1}||/||\rho_c^n||< 10^{-4}$ where
$||\rho||= \sqrt{\sum_{i}\rho(\omega_{i})^{2}}$), and
the chemical-poten\-tial self consistency is satisfied for $n_{c}$ to at
least 5 decimal places, for all fillings and temperatures.

For the case of a Bethe lattice, used in this paper, the self energy
$\Sigma_\sigma(\omega,T)$ of Eq.~(\ref{eq:sc-dmft}) can be eliminated from
the DMFT self consistency and left out of the calculation.
However, it is required to 
calculate quantities such as transport coefficients, so we outline 
here its derivation. We first separate out the $n=0$ conduction
electron orbital and re-diagonalize the remaining
orbitals, $n=1,2,\dots$, to obtain a new set of 
${\mathbf k}$-states, $a_{{\mathbf k},\sigma}$, with dispersion
$E_{{\mathbf k},\sigma}$. In terms of these, the effective 
impurity model becomes,
\begin{eqnarray}
&&{\cal H}_{imp} = J\;\sum_{\mu,\nu}{\mathbf S}\cdot
\left(b_{0,\mu}^{\dagger} {\bm \sigma}_{\mu,\nu} b_{0,\nu}\right)
+\sum_{\mu}\;\varepsilon_{0}\; 
b^{\dagger}_{0,\mu} b_{0,\mu}\nonumber\\ 
&&
+\sum_{{\mathbf k},\nu} E_{{\mathbf k},\nu}\; 
a_{{\mathbf k},\nu}^{\dagger} a_{{\mathbf k},\nu}
+\sum_{{\mathbf k},\nu}\left(V_{\mathbf k}
a_{{\mathbf k},\nu}^{\dagger} b_{0,\nu} + H.c.\right). 
\label{eq:KM-heavy-orbital}
\end{eqnarray}
The hybridization matrix elements $V_{\mathbf k}$ are introduced formally
via $\lambda_{0}\,b_{1,\sigma}=\sum_{\mathbf k}V_{\mathbf k}a_{{\mathbf k},\sigma}$,
which expresses $b_{1,\sigma}$ in terms of the new complete set of
${\mathbf k}$-states.

In order to obtain an expression for the self-energy, we follow the
method in\cite{bulla.98}, and consider equations of motion for
$G_{c,\sigma}(\omega,T)
=\langle\langle c_{0,\sigma}; c_{0,\sigma}^{\dagger}\rangle\rangle
\equiv\langle\langle b_{0,\sigma}; b_{0,\sigma}^{\dagger}
\rangle\rangle$:
\begin{eqnarray}
&&(\omega-\varepsilon_{0})G_{c,\sigma}(\omega,T) = 1 + 
\sum_{\mathbf k}V_{\mathbf k}
\langle\langle a_{{\mathbf k},\sigma};b_{0,\sigma}^{\dagger} \rangle\rangle
+ \Gamma_{\sigma}(\omega,T),\nonumber\\
&&(\omega - E_{{\mathbf k},\sigma})
\langle\langle a_{{\mathbf k},\sigma};b_{0,\sigma}^{\dagger} \rangle\rangle = 
V_{\mathbf k}G_{c,\sigma}(\omega,T),\nonumber
\end{eqnarray}
where 
\begin{equation}
\Gamma_{\sigma}(\omega,T)=
\langle\langle O_{\sigma};b_{0,\sigma}^{\dagger} \rangle\rangle \;,
\end{equation}
and
\begin{equation}
O_{\sigma}=\frac{J}{2}(b_{0,-\sigma}S^{-\sigma}+\sigma\;b_{0,\sigma}S^{z})\;.
\label{eq:operator}
\end{equation}
Solving the above equations for $G_{c,\sigma}(\omega,T)$ gives
\begin{eqnarray}
G_{c,\sigma}(\omega,T) &=& 
\left[\omega-\varepsilon_{0}-\Delta_{\sigma}(\omega,T) 
- \Sigma_{\sigma}(\omega,T)\right]^{-1}\\
\Delta_{\sigma}(\omega,T)&=&\sum_{\mathbf k}\frac{|V_{\mathbf k}|^{2}}{
(\omega-E_{\mathbf k})}\\ 
\Sigma_{\sigma}(\omega,T) &=&
\frac{\Gamma_{\sigma}(\omega,T)}{G_{c,\sigma}(\omega,T)}.
\end{eqnarray}
As discussed in Ref.~\onlinecite{bulla.98}, this form of the 
self-energy is best suited for DMFT(NRG) calculations.
The ``single-particle'' renormalization and broadening of the local 
Wannier orbital due to coupling to the remaining conduction electron
orbitals $n=1,2,\dots$ is contained in the 
hybridization function $\Delta_{\sigma}(\omega,T)$.
The many-body renormalization and broadening of this orbital
is contained in the self energy,
$\Sigma_{\sigma}$, 
which vanishes in the non-interacting limit $J\rightarrow
0$.
The DMFT self-consistency condition (\ref{eq:sc-dmft}) reduces in the
present case of a Bethe-lattice to
\begin{equation}
\varepsilon_{0}+\Delta_{\sigma}(\omega,T)=
\frac{D^{2}}{4}\;G_{c,\sigma}(\omega,T)\, .
\label{eq:sc-dmft-bl}
\end{equation}

\section{SPECTRAL FUNCTIONS}
\label{app:spectra}

The low-energy part of the $f$-electron PES of the Kondo lattice 
model (\protect\ref{eq:KLM}) is determined 
from the local T-matrix, ${\cal T}_{0}$, of (\ref{eq:KLM}), via
\begin{equation}
A(\omega,T) = -{1 \over \pi} \; {\rm Im} \; {\cal T}_{0}(\omega,T),
\label{Afunc:eq}
\end{equation}
and ${\cal T}_{0}$ is related to the ${\mathbf k}$-dependent
T-matrix of the Kondo lattice model ${\cal T}_{{\mathbf k},{\mathbf k}'}$, via
\begin{equation}
{\cal T}_{0}(\omega,T)=\sum_{{\mathbf k},{\mathbf k}'}
{\cal T}_{{\mathbf k},{\mathbf k}'}(\omega,T)\ .
\end{equation}
${\cal T}_{{\mathbf k},{\mathbf k}'}$ is defined by the identity
\begin{equation}
G_{c}^{{\mathbf k},{\mathbf k}'}=
\delta_{{\mathbf k},{\mathbf k'}}G_{0}^{{\mathbf k}{\mathbf k}}
+G_{0}^{{\mathbf k}{\mathbf k}}{\cal T}_{{\mathbf k},{\mathbf k}'}
G_{0}^{{\mathbf k}'{\mathbf k}'}.
\end{equation}
where
$G_{0}^{{\mathbf k}{\mathbf k}}$ and $G_{c}^{{\mathbf k}{\mathbf k}}$ are,
respectively,
the non-interacting and interacting conduction electron Green's functions
of the Kondo lattice model. An explicit expression 
for ${\cal T}_{0}$ has been given in Ref.~\onlinecite{costi.00}:
${\cal T}_{0} = \langle\langle O_{\sigma};O_{\sigma}^{\dagger} 
\rangle \rangle$, where the operator $O_{\sigma}$ is defined in
Eq.~\ref{eq:operator}.
Note that $A(\omega,T)$ only describes the part of the $f$-spectrum
lying close to the Fermi level and associated with the lattice 
Kondo resonance. The higher-energy $f$-electron excitations associated with
the $f$-level satellite peaks are projected out in this model.
Hence, unlike $\rho_{c}$, $A(\omega,T)$
is not normalized to unity. Its integrated weight reflects the weight
of the many-body Kondo resonance and will therefore depend strongly on 
temperature. 
Finally, for the purposes of identifying the relevant low-energy scales,
it is also useful to consider the local magnetic excitation spectrum, 
$\chi_{zz}''(\omega,T)$, of the effective quantum impurity model (\ref{eq:KM}).
This is defined by
\begin{equation}
\chi_{zz}''(\omega,T)=-\frac{1}{\pi}\; {\rm Im}\;\chi_{zz}(\omega,T),
\label{chi:eq}
\end{equation} 
where $\chi_{zz}(\omega,T)$ is the longitudinal dynamical susceptibility 
of (\ref{eq:KM}) after self-consistency is achieved.



\begin{thebibliography}{9}
\bibitem{note-T0values}
  For the values of the band filling used in our calculations we find
  $T_{0}(n_c=0.9)=1.46\times 10^{-3} D$, $T_{0}(n_c=0.8)=9.26\times 10^{-4} D$,
  $T_{0}(n_c=0.7)=5.33\times 10^{-4} D$, and $T_{0}(n_c=0.4)=9.63\times 10^{-5}
  D$.
\bibitem{note-temps}
  This natural sequence of temperatures are used for Figs.~\ref{fig3},
  \ref{fig-int-0.9}-\ref{fig-int-0.4}.  Spectra for intermediate temperatures,
  as in Figs.~\ref{fig-YbInCu4-hiT}-\ref{fig-YbAgCu4}, are obtained by
  interpolation.
\end{thebibliography}
\end{document}